\documentclass[english]{elsarticle}
\usepackage[T1]{fontenc}
\usepackage[latin9]{inputenc}
\usepackage{geometry}
\usepackage{color}
\usepackage{array}
\usepackage{float}
\usepackage{url}
\usepackage{multirow}
\usepackage{amsthm}
\usepackage{amsmath}
\usepackage{amssymb}
\usepackage{graphicx}
\usepackage{setspace}
\usepackage{esint}

\makeatletter

\providecommand{\tabularnewline}{\\}
\floatstyle{ruled}
\newfloat{algorithm}{tbp}{loa}
\providecommand{\algorithmname}{Algorithm}
\floatname{algorithm}{\protect\algorithmname}

\biboptions{semicolon}
\bibpunct{(}{)}{;}{a}{}{;}

 \title{Interpretable Whole-Brain Prediction Analysis with GraphNet}
 
 \author[mbc,stat]{Logan Grosenick\corref{cor1}\fnref{fn1}}
 \ead{logang@stanford.edu}
 
 \author[stat]{Brad Klingenberg}

 \author[psych]{Kiefer Katovich}
 
 \author[psych]{Brian Knutson}

 \author[stat]{Jonathan E. Taylor}
 
 \cortext[cor1]{Corresponding author}
 \fntext[fn1]{Correspondence to: \\ Logan Grosenick \\ 220 Panama Street, Ventura Hall Rm 30 \\ Stanford, CA 94305 \\ (650) 283-0010 }

 \address[mbc]{Center for Mind, Brain, and Computation, Stanford University, Stanford, CA, USA}
 \address[stat]{Department of Statistics, Stanford University, Stanford, CA, USA}
 \address[psych]{Deparment of Psychology, Stanford University, Stanford, CA, USA}

\makeatother

\usepackage{babel}
\begin{document}
\begin{abstract}
\begin{singlespace}
\textcolor{black}{Multivariate machine learning methods are increasingly
used to analyze neuroimaging data, often replacing more traditional
{}``mass univariate'' techniques that fit data one voxel at a time.
In the functional magnetic resonance imaging (fMRI) literature, this
has led to broad application of {}``off-the-shelf'' classification
and regression methods. These generic approaches allow investigators
to use ready-made algorithms to accurately decode perceptual, cognitive,
or behavioral states from distributed patterns of neural activity.
}However, when applied to correlated whole-brain fMRI data these methods
suffer from coefficient instability, are sensitive to outliers, and
yield dense solutions that are hard to interpret without arbitrary
thresholding.\textcolor{black}{{} Here, we develop variants of the the
Graph-constrained Elastic Net (GraphNet), a fast, whole-brain regression
and classifi{}cation method developed for spatially and temporally
correlated data that automatically yields interpretable coefficient
maps \citep{HBM2009}. GraphNet methods yield sparse but structured
solutions by combining structured graph constraints (based on knowledge
about coefficient smoothness or connectivity) with a global sparsity-inducing
prior that automatically selects important variables. Because GraphNet
methods can efficiently fit regression or classification models to
whole-brain, multiple time-point data sets and enhance classification
accuracy relative to volume-of-interest (VOI) approaches, they eliminate
the need for inherently biased VOI analyses and allow whole-brain
fitting without the multiple comparison problems that plague mass
univariate and roaming VOI ({}``searchlight'') methods. As fMRI
data are unlikely to be normally distributed, we (1) extend GraphNet
to include robust loss functions that confer insensitivity to outliers,
(2) equip them with {}``adaptive'' penalties that asymptotically
guarantee correct variable selection, and (3) develop a novel sparse
structured Support Vector GraphNet classifier (SVGN). When applied
to previously published data \citep{Knutson2007}, these efficient
whole-brain methods significantly improved classifi{}cation accuracy
over previously reported VOI-based analyses on the same data \citep{Knutson2007,Grosenick:2008p2789}
while discovering task-related regions not documented in the original
VOI approach. Critically, GraphNet estimates fit to the \citet{Knutson2007}
data generalize well to out-of-sample data collected more than three
years later on the same task but with different subjects and stimuli
\citep{Karmarkar:2012}. By enabling robust and efficient selection
of important voxels from whole-brain data taken over multiple time
points (>100,000 {}``features''), these methods enable data-driven
selection of brain areas that accurately predict single-trial behavior
within and across individuals. }
\end{singlespace}

\textcolor{black}{\pagebreak{}}
\end{abstract}
\maketitle
\begin{singlespace}

\section{\textcolor{black}{Introduction\medskip{}
}}
\end{singlespace}

\begin{singlespace}
\textcolor{black}{}
\begin{figure}
\begin{centering}
\textcolor{black}{\includegraphics[scale=0.85]{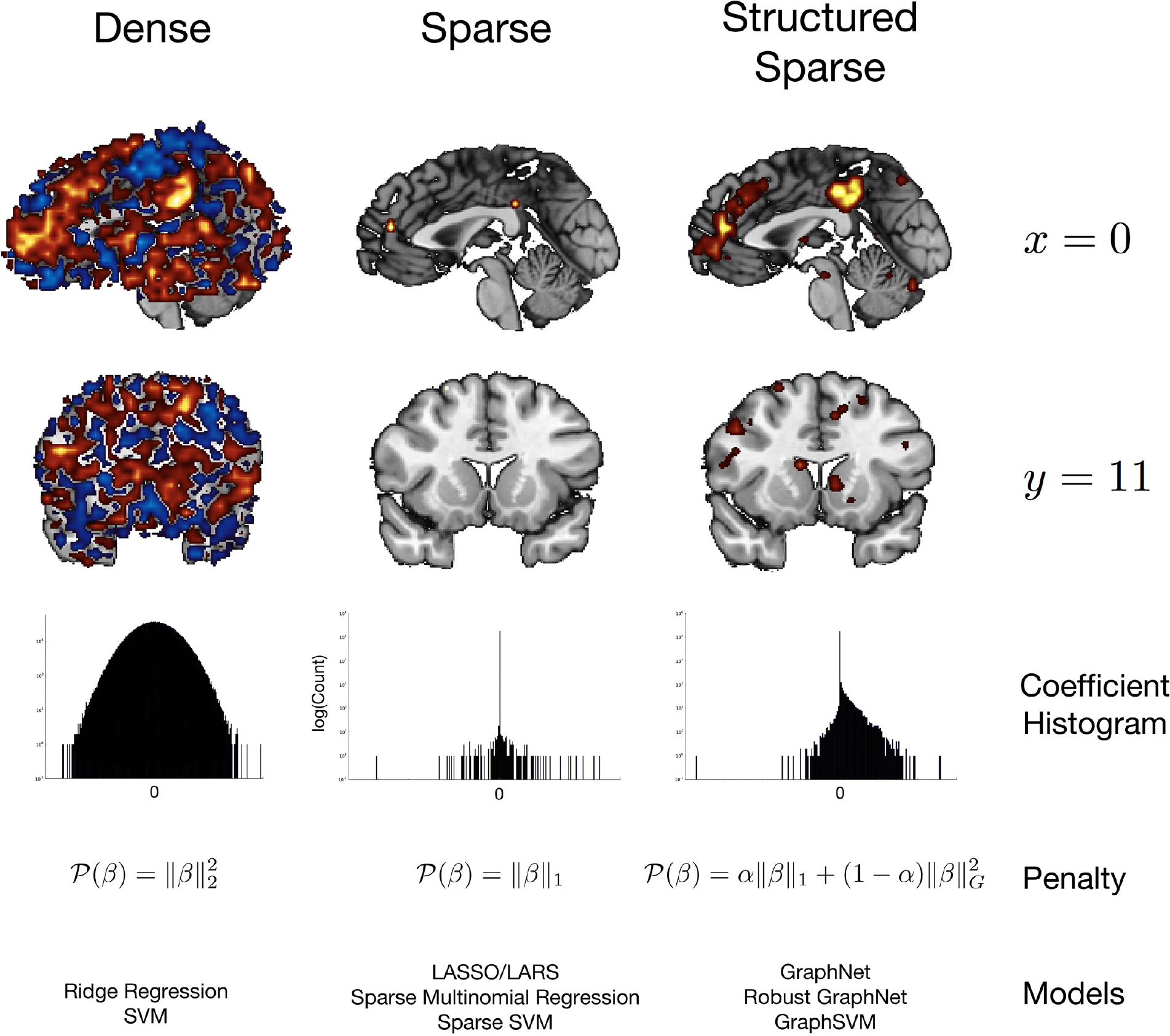}}
\par\end{centering}

\textcolor{black}{\medskip{}
}

\begin{spacing}{0.5}
\raggedright{}\textbf{\textcolor{black}{\footnotesize Figure 1. }}\textcolor{black}{\footnotesize Mid-sagittal
and coronal plots of example coefficients from dense, sparse, and
structured sparse coefficients (in Talairach coordinates). Warm colored
coefficients indicate a positive relationship with the target variable
(here predicting the decision to buy a product), cool colors a negative
relationship. Sparse methods set many  coefficients to zero, while
in dense methods almost all coefficients are nonzero. Structured sparse
methods use a penalty on differences between selected voxels to impose
a structure on the  fit so that it yields coefficients that are both
sparse and structured (e.g., smooth). Log-histograms of the estimated
voxel-wise coefficients show that the sparse method coefficients have
a near-Laplacian (double-exponential) distribution, while the dense
coefficients have a near-Gaussian distribution. The structured sparse
 coefficients are a product of these distributions (see also Figure
2). Coefficient penalties that yield each result and examples of related
methods are given below each column. }\end{spacing}
\end{figure}

\textcolor{black}{Accurately predicting subject behavior from functional
brain data is a central goal of neuroimaging research. In functional
magnetic resonance imaging (fMRI) studies, investigators measure the
blood oxygen-level dependent (BOLD) signal---a proxy for neural activity---and
relate this signal to psychophysical or psychological variables of
interest. Historically, modeling is performed one voxel at a time
to yield a map of univariate statistics that are then thresholded
according to some heuristic to yield a {}``brain map'' suitable
for visual inspection. Over the past decade, however, a growing number
of neuroimaging studies have applied machine learning analyses to
fMRI data to model effects across multiple voxels. Commonly referred
to as {}``multivariate pattern analysis'' \citep{Hanke:2009p615}
or {}``decoding'' (to distinguish them from more commonly-used {}``mass-univariate''
methods \citep{Friston:1995p2108}), these approaches have allowed
investigators to use activity patterns across multiple voxels to classify
image categories during visual presentation \citep{Peelen:2009p2210,Shinkareva:2008p2395},
image categories during memory retrieval \citep{Polyn:2005p2405},
intentions to move \citep{Haynes:2007p2478}, and even intentions
to purchase \citep{Grosenick:2008p2789} (to name just a few applications---see
also \citet{Norman:2006p824,Haynes:2006p862,Pereira:2009p606,OToole:2007p909,Bray2009},
and examples in }\textit{\textcolor{black}{NeuroImage}}\textcolor{black}{{}
Volume 56 Issue 2). In multiple cases, these statistical learning
algorithms have shown better predictive performance than standard
mass-univariate analyses \citep{Haynes:2006p862,Pereira:2009p606}.}

\textcolor{black}{Despite these advances, analysis of neuroimaging
data with statistical learning algorithms is still young. Most of
the research that has applied statistical learning algorithms to fMRI
data has been conducted by a few laboratories \citep{Norman:2006p824},
and most analyses have been conducted with off-the-shelf classifiers
(\citet{Norman:2006p824,Pereira:2009p606}, but cf. \citealt{Grosenick:2008p2789,Hutchinson:2009p2394,Chappell2009,Brodersen2011,Michel2011,NgVaroquaux2012}).
These classifiers are often applied to volume of interest (VOI) data
within subjects rather than whole-brain data across subjects (\citet{Etzel:2009p3221,Pereira:2009p606},
but cf. \citealt{Mitchell:2004p970,MouraoMiranda:2007p2565,HBM2009,HBM2010,Ryali2010,VanGervenHeskes2012,Michel2011,NgVaroquaux2012}).
While these classifiers have a venerable history in the machine learning
literature, they were not originally developed for application to
whole-brain neuroimaging data, and so suffer from inefficiencies in
this context. Specifically, the large number of features (usually
voxel data) and spatiotemporal correlations characteristic of fMRI
data present unique challenges for off-the-shelf classifiers.}

\textcolor{black}{Indeed, the purpose of off-the-shelf classifiers
in the machine learning literature (e.g., discriminant analysis (DA),
naive Bayes (NB), k-nearest neighbors (kNN), random forests (RF),
and support vector machines (SVM)) has been to quickly and easily
yield good classification accuracy---for example in example speech
recognition or hand-written digit identification \citep{Hastie:2009p2681}.
Beyond accuracy, however, neuroscientists often aim to understand
which neural features are related to particular stimuli or behaviors
at specific points in time. This distinct aim of interpretability
requires classification or regression methods that can yield clearly
interpretable sets of model coefficients. For this reason, the recent
literature on classification of fMRI data has recommended using linear
classifiers (e.g., logistic regression (LR), linear discriminant analysis
(LDA), Gaussian Naive Bayes (GNB), or linear SVM) rather than nonlinear
classifiers \citep{Haynes:2006p862,Pereira:2009p606}. }

\textcolor{black}{Linearity alone, however, does not guarantee that
a method will yield a stable and interpretable solution. For instance,
in the case of multiple correlated input variables LR, LDA, and GNB
yield unstable coefficients and degenerate covariance estimates, particularly
when applied to smoothed data \citep{Hastie:1995p2589,Hastie:2009p2681}.
In the context of classification, penalized least squares may over
smooth coefficients, complicating interpretation \citep{Friedman1997}.
Additionally, most linear classifiers return dense sets of coefficients
(as in Figure 1, left panels) that require subsequent thresholding
or feature selection to yield parsimonious solutions. Although heuristic
methods exist for coefficient selection, these are generally greedy
(e.g., forward/backward stage-wise procedures like Recursive Feature
Elimination \citep{Guyon2002,DeMartino2007,Bray2009}), yielding unstable
solutions when data are resampled (since these algorithms tend to
converge to local minima) \citep{Hastie:2009p2681}. Although principled
methods exist for applying thresholds to dense mass-univariate coefficient
maps (e.g. Random Field Theory \citep{AdlerTaylor2000,Worsley2004}),
these approaches do not currently extend to dense multivariate regression
or classification methods. }

\textcolor{black}{Recently, sparse regression methods have been applied
to neuroimaging data to yield reduced coefficient sets that are automatically
selected during model fitting. The first examples in the fMRI literature
include \citet{Yamashitaa2008}, who applied sparse logistic regression
\citep{Tibs1996} to classification of visual stimuli, and \citet{Grosenick:2008p2789}
who first developed sparse penalized discriminant analysis by converting
an {}``Elastic Net'' regression \citep{ZouHastie} into a classifier,
and then applied it to choice prediction. Subsequently, sparse methods
for regression \citep{Carroll:2009p2920,Hanke:2009p615} and classification
\citep{Hanke:2009p615} have been applied to fMRI data to yield reduced
sets of coefficients from volumes of interest, whole-brain volumes
\citep{Ryali2010,vanGerven2010}, and whole-brain volumes over multiple
time points \citep{HBM2009,HBM2010}. These methods typically impose
an $\ell_{1}$-penalty (sum of absolute values) on the model coefficients,
which sets many of the estimated coefficients to zero (see Figure
1, leftmost panels, and Figure 2b). When applied to correlated fMRI
data, however, $\ell_{1}$-penalized methods can select an overly
sparse solution--resulting in omission of relevant features as well
as unstable coefficient estimates during cross-validation \citep{ZouHastie,Grosenick:2008p2789}.
To allow relevant but correlated coefficients to coexist in a sparse
model fit, recent approaches to fMRI regression \citep{Carroll:2009p2920,Li:2009p2898}
and classification \citep{Grosenick:2008p2789,HBM2009,Ryali2010}
impose a hybrid of both $\ell_{1}$- and $\ell_{2}$-norm penalties
(the {}``Elastic Net'' penalty of \citet{ZouHastie}) on the coefficients.
These hybrid approaches allow the inclusion of correlated variables
in sparse model fits. }

\textcolor{black}{This paper explores modified methods that combine
the Elastic Net penalty with a general user-specified sparse graph
penalty. This sparse graph penalty allows the user to efficiently
incorporate physiological constraints and prior information (such
as smoothness in space or time or anatomical details such as topology
or connectivity) in the model. The resulting graph-constrained elastic
net (or {}``GraphNet'') regression \citep{HBM2009,HBM2010} has
the capacity to find {}``structured sparsity'' in correlated data
with many features (Figure 1, right panels), consistent with results
in the manifold learning \citep{Belkin2006} and gene microarray literatures
\citep{LiLi2008}. In the statistics literature, related {}``sparse
structured'' methods have been shown to have desirable convergence
and variable selection properties for large correlated data sets \citep{SlawskiTutz2010,Jenatton2011}.
These sparse, structured models can also be implemented within a Bayesian
framework \citep{vanGerven2010}. Here, we extend the performance
of GraphNet regression and classification methods to whole-brain fMRI
data by: (1) generalizing them to be robust to outliers in fMRI data
(for both regression and classification), (2) adding {}``adaptive''
penalization to reduce fit bias and improve variable selection, and
(3) developing a novel support vector GraphNet (SVGN) classifier.
Additionally, to efficiently fit GraphNet methods to whole-brain fMRI
data over multiple time-points, we adapt algorithms from the applied
statistics literature \citep{Friedman2010}. }

\textcolor{black}{After developing robust and adaptive GraphNet regression
and classification methods, we demonstrate the enhanced performance
of GraphNet classifiers on previously published data \citep{Knutson2007,Karmarkar:2012}.
Specifically, we use GraphNet methods to predict subjects' trial-to-trial
purchasing behavior with whole-brain data over several time points,
and then infer which brain regions best predict upcoming choices to
purchase or not purchase a product. Fitting these methods to 25 subjects'
whole-brain data over 7 time points (2s TRs) yielded classification
rates which exceeded those found previously in a volume of interest
(VOI) based classification analysis \citep{Grosenick:2008p2789},
as well as those obtained with a linear support vector machine (SVM)
classifier fit to the whole brain data. While the GraphNet results
on whole-brain data confirm the relevance of previously chosen volumes
of interest (i.e., bilateral nucleus accumbens (NAcc), medial prefrontal
cortex (MPFC), and anterior insula), they also implicate previously
unchosen areas (i.e., ventral tegmental area (VTA) and posterior cingulate).
We conclude with a discussion of the interpretation of GraphNet model
coefficients, as well as future improvements, applications, and extensions
of this family of GraphNet methods to neuroimaging data. Open source
code for solving the GraphNet problems in this paper is freely available
at \url{https://github.com/logang/neuroparser}.}
\end{singlespace}

\textcolor{black}{\medskip{}
}

\begin{singlespace}

\section{\textcolor{black}{Methods}}
\end{singlespace}

\subsection{Background\textcolor{black}{\medskip{}
}}

\subsubsection{\textcolor{black}{Penalized least squares }}

\textcolor{black}{\medskip{}
}

\begin{singlespace}
\textcolor{black}{Many classification and regression problems can
be formulated as modeling a response vector $y\in\mathbb{R}^{n}$
as a function of data matrix $X\in\mathbb{R}^{n\times p}$, which
consists of $n$ observations each of length $p$ (with $n\geq p)$.
In particular, a large number of models treat $y$ as a linear combination
of the predictors in the presence of noise $\epsilon\in\mathbb{R}^{n}$,
such that}

\textcolor{black}{
\begin{equation}
y=X\beta+\epsilon,\label{eq:1}
\end{equation}
where $\epsilon$ is a noise vector typically assumed to be normally
distributed $\epsilon\sim\mathcal{N}(0,I\sigma^{2})$ with vector
mean $0$ and diagonal variance-covariance matrix $I\sigma^{2}$,
and $\beta\in\mathbb{R}^{p}$ a vector of linear model coefficients.
In this case using squared error loss leads to the well-known ordinary
least squares (OLS) solution}

\textcolor{black}{
\begin{equation}
\widehat{\beta}=\underset{\beta}{\text{argmin}}\ \|y-X\beta\|_{2}^{2}=(X^{T}X)^{-1}X^{T}y,\label{eq:2}
\end{equation}
which yields the best linear unbiased estimator (BLUE) if the columns
of $X$ are uncorrelated \citep{LehmannCasella}. }

\textcolor{black}{However, this estimator is inefficient in general
for}\textbf{\textcolor{black}{{} }}\textcolor{black}{$p>2$---it is
dominated by biased estimators \citep{Stein1955}---and if the columns
of $X$ are correlated (i.e. are {}``multicollinear'') then the
estimated coefficient values can vary erratically with small changes
in the data, so the OLS fit can be quite poor. A common solution to
this problem is penalized (or {}``regularized'') least squares regression
\citep{Tikhonov1943}, in which the magnitude of the model coefficients
are penalized to stabilize them. This is accomplished by adding a
penalty term $\mathcal{P}(\beta)$ on the coefficient vector $\beta$,
yielding}

\textcolor{black}{
\begin{equation}
\widehat{\beta}=\underset{\beta}{\text{argmin}}\ \|y-X\beta\|_{2}^{2}+\lambda\mathcal{P}(\beta),\ \lambda\in\mathbb{R}_{+},\label{eq:3}
\end{equation}
where $\lambda$ is a parameter that trades off least squares goodness-of-fit
with the penalty on the model coefficients (or equivalently, trades
off fit variance for fit bias) and $\mathbb{R}_{+}$ is the set of
nonnegative scalars. These estimates are equivalent to maximum a posteriori
(MAP) estimates from a Bayesian perspective (with a Gaussian prior
on the coefficients if $\mathcal{P}(\beta)=\|\beta\|_{2}^{2}$ \citep{Hastie:2009p2681}),
or to the Lagrangian relaxation of a constrained bi-criterion optimization
problem \citep{BoydVandenberghe2004}. Such equivalencies motivate
various interpretations of the model coefficients and parameter $\lambda$
(see section 2.3). \medskip{}
}
\end{singlespace}

\begin{singlespace}

\subsubsection{\textcolor{black}{Sparse regression and automatic variable selection }}
\end{singlespace}

\textcolor{black}{\medskip{}
}

\begin{singlespace}
\textcolor{black}{}
\begin{figure}[t]
\begin{centering}
\textcolor{black}{\includegraphics[scale=0.65]{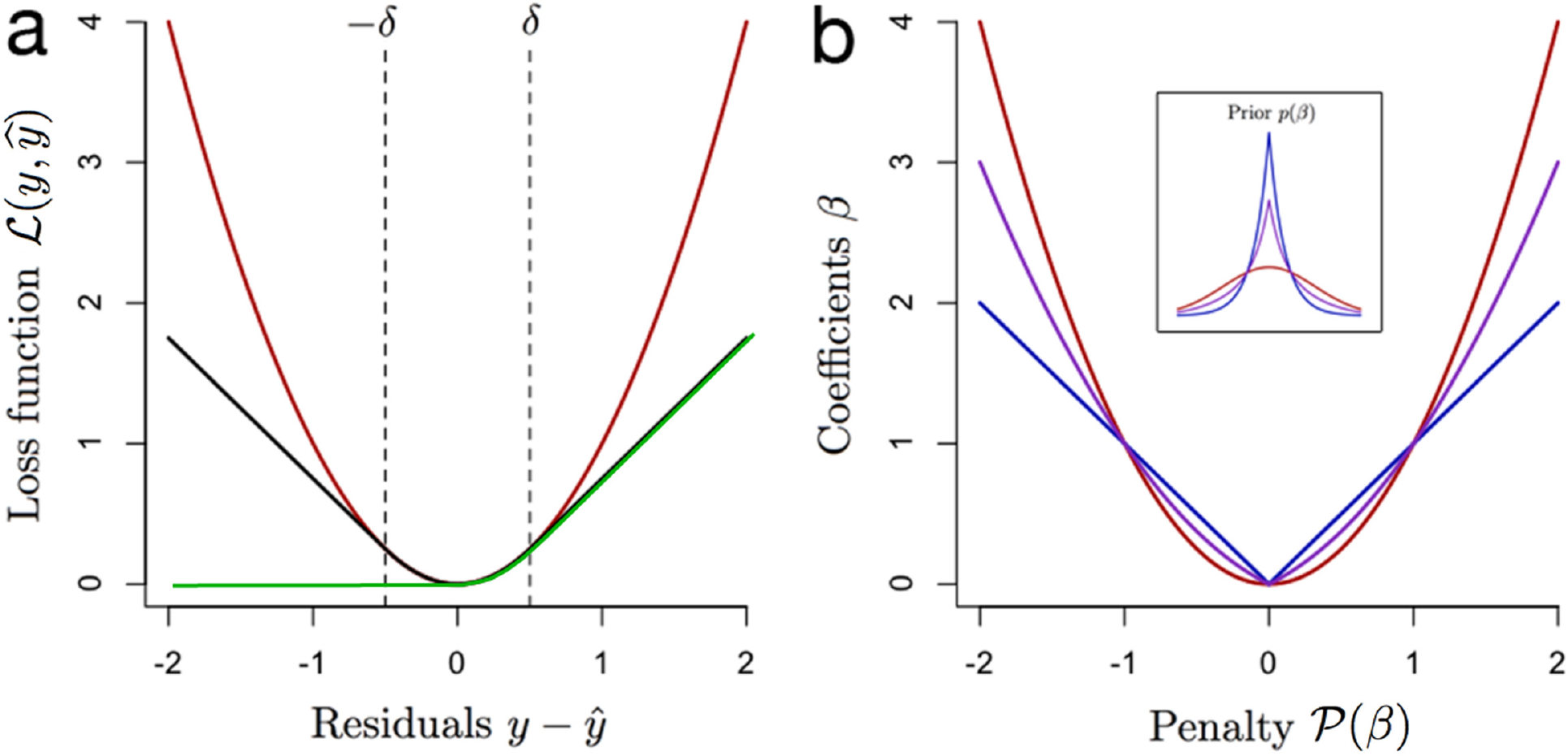}}
\par\end{centering}

\textcolor{black}{\medskip{}
}

\begin{spacing}{0.5}
\raggedright{}\textbf{\textcolor{black}{\footnotesize Figure 2. }}\textcolor{black}{\footnotesize (a)
Diagrammatic representation of squared-error (red), Huber (black),
and Huberized Hinge (green) loss functions. Dotted lines denote where
the Huber loss changes from penalizing residuals quadratically (where
$|y-\widehat{y}|\leq\delta$) to penalizing them linearly (where $|y-\widehat{y}|>\delta$).
The linear penalty on large residuals makes the Huber loss robust.
(b) Diagrammatic representation of convex penalty functions used in
this article (along one coordinate $\beta$). The red curve is a quadratic
penalty $\mathcal{P}(\beta)=\beta^{2}$ on coefficient magnitude,
often called the Tikhonov or {}``ridge'' penalty in regression.
The blue curve is the Lasso penalty on coefficient magnitude $\mathcal{P}(\beta)=|\beta|$.
The purple curve is a convex combination of the red and blue curves:
$\mathcal{P}(\beta)=\alpha\beta^{2}+(1-\alpha)|\beta|$ (where here
$\alpha=0.5)$, called the {}``Elastic Net'' penalty. The inset
shows the shape of the prior distribution on the coefficient estimates
that each of these penalties corresponds to: Gaussian (red), Laplacian
(blue), and mixed Gaussian and Laplacian (purple) (units arbitrary).
The priors become increasingly peaked around zero as the Elastic Net
penalty approaches the Lasso penalty, corresponding to a prior belief
that many coefficients will be exactly zero.}\end{spacing}
\end{figure}
\textcolor{black}{{} There are a few standard choices for the penalty
$\mathcal{P}(\beta)$. Letting $\mathcal{P}(\beta)=\|\beta\|_{2}^{2}=\sum_{j=1}^{p}\beta_{j}^{2}$
(the $\ell_{2}$ norm) gives the classical Tikhonov or {}``ridge''
regression estimates originally proposed for such problems \citep{Tikhonov1943,HoerlKennard1970}.
More recently, the choice $\mathcal{P}(\beta)=\|\beta\|_{1}=\sum_{j=1}^{p}|\beta_{j}|$
(the $\ell_{1}$ norm)---called the Least Absolute Shrinkage and Selection
Operator (or {}``Lasso'') penalty in the regression context \citep{Tibs1996}---has
become widely popular in statistics, engineering, and computer science,
leading some to call such $\ell_{1}$-regression the {}``modern least
squares'' \citep{CandesBoyd2008}. In addition to shrinking the coefficient
estimates, the Lasso performs variable selection by producing sparse
coefficient estimates (i.e., many are exactly equal to zero, see Figure
1 left panels). In many applications, having a sparse vector $\widehat{\beta}$
is highly desirable, since a fit with fewer non-zero coefficients
is simpler, and can help select predictors that have an important
relationship with the response variable $y$. }

\textcolor{black}{The $\ell_{1}$-norm used in the Lasso is the closest
convex relaxation of the $\ell_{0}$ pseudo-norm $\|\beta\|_{0}=\sum_{j=1}^{p}1_{\{\beta_{j}\ne0\}}$,
where $1_{\{\beta_{j}\ne0\}}$ is an indicator function that is $1$
if the $j$th coefficient $\beta_{j}$ is nonzero and $0$ otherwise.
This represents a penalty on the number of nonzero coefficients (their
cardinality). However, finding a minimal cardinality solution generally
involves a combinatorial search through possible sets of nonzero coefficients
(a form of {}``all subsets regression'' \citep{Hastie:2009p2681})
and so is computationally infeasible for even a modest number of input
features. An $\ell_{1}$-norm penalty can be used as a heuristic that
results in coefficient sparsity (which corresponds to the maximum
a posteriori (MAP) estimates under a Laplacian (double-exponential)
prior; for a fully Bayesian approach see \citet{vanGerven2010}).
Such $\ell_{1}$-penalized regression methods set many variables equal
to zero and automatically select only a small subset of relevant variables
to assign nonzero coefficients. While these methods yield the sparsest
possible fit in many cases \citep{DonohoElad2003,Donoho2006}, they
do not always do so, and reweighted methods (e.g., Automatic Relevance
Determination (ARD) \citep{Wipf2008} and iterative reweighting of
the $\ell_{1}$ penalty \citep{CandesBoyd2008}) exist for finding
sparser solutions. It is worth noting that while Bayesian methods
for variable selection (such as Relevance Vector Machines) have existed
in the literature for some time, these methods typically require using
EM-like or MCMC approaches that do not guarantee convergence to a
global minimum and that are relatively computationally inefficient
(}though see \citet{MohamedHeller2011} for an interesting counter-point\textcolor{black}{).
As an interesting exception, recent work on ARD and sparse Bayesian
learning \citep{Wipf2008} has provided an attractive alternative,
showing that the sparse Bayesian learning problem can be solved as
a sequence of reweighted Lasso problems, similar to the adaptive methods
discussed below. This approach no longer provides a full posterior,
but does provide an interesting and computationally tractable link
to the Bayesian formulation. In the future we expect that such links
will lead to better approaches for model selection in these methods
than the {}``brute force'' grid search employed here. \medskip{}
}
\end{singlespace}

\subsubsection{Elastic Net regression}

\begin{singlespace}
\textcolor{black}{\medskip{}
}

\textcolor{black}{Despite offering a sparse solution and automatic
variable selection, there are several disadvantages to using $\ell_{1}-$penalized
methods like the Lasso in practice. For example, from a group of highly
correlated predictors, the Lasso will typically select a subset of
{}``representative'' predictors to include in the model fit \citep{ZouHastie}.
This can make it difficult to interpret coefficients because those
that are set to $0$ may still be useful for modeling $y$ (i.e.,
false negatives are likely). Worse, entirely different subsets of
coefficients may be selected when the data are resampled (e.g., during
cross-validation). Moreover, the Lasso can select at most $n$ non-zero
coefficients \citep{ZouHastie}, which may prove undesirable when
the number of input features ($p$) exceeds the number of observations
($n$) (i.e., {}``$p\gg n$'' problems). Finally, as a global shrinkage
method, the Lasso biases model coefficients towards zero \citep{Tibs1996,Hastie:2009p2681},
making interpretation with respect to original data units difficult.
Other methods that use only an $\ell_{1}$ penalty (e.g., sparse logistic/multinomial
regression and sparse SVM \citep{Hastie:2009p2681}) are subject to
the same deficiencies. }

\textcolor{black}{In response to several of these concerns \citet{ZouHastie}
proposed the Elastic Net, which uses a mixture of $\ell_{1}$- and
$\ell_{2}$-norm regularization, and may be written }

\textcolor{black}{
\begin{equation}
\widehat{\beta}=\kappa\ \underset{\beta}{\text{argmin}}\ \|y-X\beta\|_{2}^{2}+\lambda_{1}\|\beta\|_{1}+\lambda_{2}\|\beta\|_{2}^{2},\label{eq:4}
\end{equation}
where the factor $\kappa=1+\lambda_{2}$ in \eqref{eq:4} and subsequent
equations is a rescaling factor discussed in further detail below.
This Elastic Net estimator overcomes several (though not all) of the
disadvantages discussed above, while maintaining many advantages of
Tikhonov ({}``Ridge'') regression and the Lasso. In particular,
the Elastic Net accommodates groups of correlated variables and can
select up to $p$ variables with non-zero coefficients. The amount
of sparsity in the solution vector can be tuned by adjusting the penalty
coefficients $\lambda_{1}$ and $\lambda_{2}$. In this case, the
$\ell_{1}$ penalty can be thought of as a heuristic for enforcing
sparsity, while the $\ell_{2}$ penalty allows correlated variables
to enter the model and stabilizes the sample covariance estimate.
This Elastic Net approach performs well on fMRI data in both regression
and classification settings \citep{Grosenick:2008p2789,Carroll:2009p2920,Ryali2010}.\medskip{}
}
\end{singlespace}

\begin{singlespace}

\subsubsection{\textcolor{black}{Graph-constrained Elastic Net (GraphNet) regression }}
\end{singlespace}

\textcolor{black}{\medskip{}
}

\begin{singlespace}
\textcolor{black}{So far we have seen that sparse regression methods
like the Elastic Net, which use a hybrid $\ell_{1}$- and $\ell_{2}$-norm
penalty, can be used to yield sparse model fits that do not exclude
correlated variables \citep{ZouHastie}, and that we can turn these
regression methods into classifiers that perform well when fit to
VOI data \citep{Grosenick:2008p2789}. However, the Elastic Net penalty
merely makes the model fitting procedure {}``blind'' to correlations
between input features (by shrinking the sample estimate of the covariance
matrix towards the identity matrix). Indeed, if $\lambda_{2}$ in
equation \eqref{eq:5} grows large, this method is equivalent to applying
a threshold to mass-univariate OLS regression coefficients (i.e.,
the estimate of the covariance matrix becomes a scaled identity matrix)
\citep{ZouHastie}.}

\textcolor{black}{In this section, we describe a modification of the
Elastic Net that explicitly imposes structure on the model coefficients.
This allows the analyst to pre-specify constraints on the model coefficients
(e.g., based on prior information like local smoothness or connectivity,
or other desirable fit properties), and then to tune how strongly
the fit adheres to these constraints. Since the user-specified constraints
take the general form of an undirected graph, we call this regression
method the graph-constrained Elastic Net (or {}``GraphNet'') \citep{HBM2009,HBM2010}. }

\textcolor{black}{The GraphNet model closely resembles the Elastic
Net model, but with a modification to the $\ell_{2}$-norm penalty
term: }

\textcolor{black}{
\begin{eqnarray}
\widehat{\beta} & = & \kappa\ \underset{\beta}{\text{argmin}}\ \|y-X\beta\|_{2}^{2}+\lambda_{1}\|\beta\|_{1}+\lambda_{G}\|\beta\|_{G}^{2}\label{eq:8}\\
\text{} &  & \|\beta\|_{G}^{2}=\beta^{T}G\beta=\sum_{j=1}^{p}\sum_{k=1}^{p}\beta_{j}G_{jk}\beta_{k},\nonumber 
\end{eqnarray}
where $G$ is a sparse graph. Note that in the case where $G=I$,
where $I$ denotes the identity matrix, the GraphNet reduces back
to the Elastic Net. Thus the Elastic Net is a special case of GraphNet
and we can replicate the effects of increasing an Elastic Net penalty
by adding a scaled version of the identity matrix $(\lambda_{2}/\lambda_{G})I$
to $G$ (for $\lambda_{G}>0$). }

\textcolor{black}{The example we will use for the matrix $G$ in the
remainder of this paper is the graph Laplacian, which formalized our
intuition that voxels that are neighbors in time and space should
typically have similar values. If we take the coefficients $\beta$
to be functions over the brain volume $V\in\mathbb{R}{}^{3}$ over
time points $T\in\mathbb{R}$ such that $\beta(x,y,z,t)$, then we
would like a penalty that penalizes roughness in the coefficients
as measured by their derivatives over space and time, such as 
\begin{equation}
\mathcal{P}(\beta)=\int_{V,T}\left(\frac{\partial^{2}\beta}{\partial x^{2}}+\frac{\partial^{2}\beta}{\partial y^{2}}+\frac{\partial^{2}\beta}{\partial z^{2}}+\frac{\partial^{2}\beta}{\partial t^{2}}\right)dx\ dy\ dz\ dt=\int_{V,T}\Delta\beta\ dx\ dy\ dz\ dt,\label{eq:9}
\end{equation}
where $\Delta$ is the Laplacian operator, which here is a 4D isotropic
measure of the second spatio-temporal derivative of the volumetric
time-series. Since we are sampling discretely, we use the discrete
approximation to the Laplacian operator $\Delta$: the matrix Laplacian
$L=D-A$ (the difference between the degree matrix $D$ and the adjacency
matrix $A$, see e.g., \citep{Hastie:1995p2589}). This formulation
generalizes well to arbitrary graph connectivity and is widely used
in spectral clustering techniques and spectral graph theory \citep{BelkinNiyogi2008}. }

\textcolor{black}{In the case where $G=L$, the graph penalty, $\|\beta\|_{G}^{2}$,
has the appealingly simple representation
\[
\|\beta\|_{G}^{2}=\sum_{(i,j)\in\mathcal{E}_{G}}(\beta_{i}-\beta_{j})^{2},
\]
where $\mathcal{E}_{G}$ is the set of index pairs for voxels that
share an edge in graph $G$ (i.e. have a nonzero entry in the adjacency
matrix $A$). Written this way, the graph penalty induces smoothness
by penalizing the size of the pairwise differences between coefficients
that are adjacent in the graph. In the one dimensional case, if the
quadratic terms $(\beta_{i}-\beta_{j})^{2}$ were replaced by absolute
deviations $|\beta_{i}-\beta_{j}|$ then this would instead be an
instance of the \textquotedbl{}fused Lasso\textquotedbl{} \citep{Tibshirani2005}
or Generalized Lasso \citep{TibshiraniTaylor2012}. There are two
main reasons for preferring a quadratic penalty in the present application:}
\end{singlespace}
\begin{enumerate}
\begin{singlespace}
\item \textcolor{black}{The fused Lasso is closely related to Total Variation
(TV) denoising \citep{Rudin1992} and tends to set many of the pairwise
differences $\beta_{i}-\beta_{j}$ to zero, creating a sharp piecewise
constant set of coefficients that lacks the spatial smoothness often
expected in fMRI data. Extending this formulation to processes with
more that one spatial or temporal dimension is nontrivial \citep{Michel2011}.}\end{singlespace}

\item \textcolor{black}{Significant algorithmic complications can be avoided
by choosing a differentiable penalty on the pairwise differences \citep{Tseng2001,Friedman:2007p36},
speeding up model fitting and reducing model complexity considerably---especially
in the case of spatial data, where the Total Variation penalty must
be formulated as a more complicated sum of non-smooth norms on each
of the first-order forward finite difference matrices \citep{Wang2008_TV,Michel2011}.}
\end{enumerate}
Thus GraphNet methods provide a sparse and structured solution similar
to the Fused Lasso, Generalized Lasso, and Total Variation. However,
unlike these approaches, GraphNet methods allow for smooth rather
than piecewise constant structure in the non-sparse parts of the reconstructed
volume. This is of interest in cases where we might expect the magnitude
of nonzero coefficients to be different within a volume of interest.
Due to the smoothness of the graph penalty GraphNet methods are also
easier from an optimization perspective. Of course, there are certainly
situations in which the piecewise smoothness of Total Variation could
be a better prior (this depends on the data and problem formulation).\medskip{}

\begin{singlespace}

\subsubsection{\textcolor{black}{Adaptive GraphNet regression}}
\end{singlespace}

\textcolor{black}{\medskip{}
}

\begin{singlespace}
\textcolor{black}{The methods described above automatically select
variables by shrinking coefficient estimates towards zero, resulting
in downwardly biased coefficient magnitudes. This shrinkage makes
it difficult to interpret coefficient magnitude in terms of original
data units, and severely restricts the conditions under which the
Lasso can perform consistent variable selection \citep{Zou2006}.
Ideally, given infinite data, the method would select the correct
parsimonious set of features (i.e., the {}``true model'', were it
known), but avoid shrinking nonzero coefficients that remain in the
model (unbiased estimation). Together, these desiderata are known
as the {}``oracle'' property \citep{FanLi}. Note that in the neuroimaging
context, the first (consistent variable selection) corresponds to
correct localization of signal, while the second (consistent coefficient
estimation) relates to improving estimates of coefficient magnitude.}

\textcolor{black}{Several estimators possessing the oracle property
(given certain conditions on the data) have been reported in the literature,
including the Adaptive Lasso \citep{Zou2006,Zhou:2009p3000} and the
Adaptive Elastic Net \citep{Zou:2009p2991}. These estimators are
straightforward modifications of penalized linear models. They work
by starting with some initial estimates of the coefficients obtained
by fitting the non-adaptive model \citep{Zou:2009p2991}, and use
these to adaptively reweight the penalty on each individual coefficient
$\beta_{j},\ j=1,\ldots,p$. Recently \citep{SlawskiTutz2010} extended
the adaptive approach to a sparse, structure method equivalent to
GraphNet regression, and proved that the oracle properties previously
shown for the adaptive Lasso and Adaptive Elastic Net extend to the
sparse, structured case provided the true coefficients are in the
null space of $G$ (i.e. the nonzero entries of $\beta$ specify a
connected component in $G$). We refer the reader to \citep{SlawskiTutz2010}
for further details.}
\end{singlespace}

\textcolor{black}{As in \citep{SlawskiTutz2010}, we may rewrite the
GraphNet to have an adaptive penalty (the adaptive GraphNet) as follows:}

\begin{singlespace}
\textcolor{black}{
\begin{eqnarray}
\widehat{\beta} & = & \underset{\beta}{\text{argmin}}\ \|y-X\beta\|_{2}^{2}+\lambda_{1}^{*}\sum_{j=1}^{p}\widehat{w}_{j}|\beta_{j}|+\lambda_{G}||\beta||_{G}^{2}\label{eq:12}\\
\widehat{w}_{j} & = & \left|\tilde{\beta}_{j}\right|^{-\gamma}.\label{eq:reweights}
\end{eqnarray}
The idea here is that important coefficients will have large starting
estimates $\tilde{\beta}_{j}$ (where $\tilde{\beta_{j}}$ is a suitable
estimator of $\beta_{j}$) and so will be shrunk at a rate inversely
proportional to their starting estimates, leaving them asymptotically
unbiased. On the other hand, coefficients with small starting estimates
$\tilde{\beta}_{j}$ will experience additional shrinkage, making
them more likely to be excluded. We let $\gamma=1$ as in the finite
sample case \citep{Zou2006,Zou:2009p2991}, and by analogy to the
Adaptive Elastic Net \citep{Zou:2009p2991} set $\tilde{\beta}$ to
the standard GraphNet coefficient estimates for a fixed value of $\lambda_{G}$
(chosen based on the GraphNet performance at that value). We use $\lambda_{1}^{*}$
to differentiate the adaptive fit sparsity parameter from the parameter
associated with the GraphNet fit used to initialize the weights $\widehat{w}_{j}.$}
\end{singlespace}

\textcolor{black}{It is important to note that the oracle properties
that hold in the asymptotic case may not apply to the finite sample,
$p\gg n$ situation. Nevertheless, we include these methods for comparison
since oracle properties are desirable and since evidence suggests
that the adaptive elastic-net has improved fi{}nite sample performance
because it deals well with collinearity \citep{Zou:2009p2991}.}

\textcolor{black}{\medskip{}
}

\begin{singlespace}

\subsubsection{\textcolor{black}{Turning sparse regression methods into classifiers:
Optimal Scoring (OS) and Sparse Penalized Discriminant Analysis (SPDA)}}
\end{singlespace}

\textcolor{black}{\medskip{}
}

\begin{singlespace}
\textcolor{black}{Sparse regression methods like the Lasso or Elastic
Net can be turned into sparse classifiers \citep{Leng2008,Grosenick:2008p2789,Clemmensen2011}.
Naively, we might imagine performing a two-class classification simply
by running a regression with Lasso or the Elastic Net on a target
vector containing $1$'s and 0's depending on the class of each observation
$y_{i}\in\{0,1\}$. We would then take the predicted values from the
regression $\widehat{y}$ and classify to $0$ if the $i$th estimate
$\widehat{y}_{i}<0.5$ and to $1$ if the estimate $\widehat{y}_{i}>0.5$
(for example). In the multi-class case (i.e. $J$ classes with $J>2$),
multi-response linear regression could be used as a classifier in
a similar way. This would be done by constructing an indicator response
matrix $Y$, with $n$ rows and $J$ columns (where again $n$ is
the number of observations and $J$ is the number of classes). Then
the $i$th row of $Y$ has a $1$ in the $j$th column if the observation
is in the $j$th class and a $0$ otherwise. If we run a multiple
linear regression of $Y$ on predictors $X$, we can classify by assigning
the $i$th observation to the class having the largest fitted value
$\widehat{Y}_{i1},\widehat{Y}_{i2},...,\widehat{Y}_{iJ}$. With the
exception of binary classification on balanced data, this classifier
has several disadvantages. For instance, the estimates $\widehat{Y}_{ij}$
are not probabilities, and in the multi-class case certain classes
can be {}``masked'' by others, resulting in decreased classification
accuracy \citep{Hastie:2009p2681}. However, applying LDA to the fitted
values of such a multiple linear regression classifier is mathematically
equivalent to fitting the full LDA model \citep{BriemanIhaka1984},
yielding posterior probabilities for the classes and dramatically
improving classifier performance over the original multivariate regression
in some cases \citep{Hastie:1994p2648,Hastie:1995p2589,Hastie:2009p2681}. }

\textcolor{black}{\citet{Hastie:1994p2648} and \citet{Hastie:1995p2589}
exploit equivalences between multiple regression and LDA and between
LDA and canonical correlation analysis to develop a procedure they
call Optimal Scoring (OS). OS allows us to build a classifier by first
fitting a multiple regression to $Y$ using an arbitrary regression
method, and then linearly transforming the fitted results of this
regression using the OS procedure (see }\citet{Hastie:1994p2648}\textcolor{black}{{}
for further algorithmic and statistical details). This procedure yields
both class probability estimates and discriminant coordinates, and
allows us to use any number of regression methods as discriminant
classifiers. This approach is discussed in detail for nonlinear regression
methods applied to a few input features in \citet{Hastie:1994p2648},
and for regularized regression methods applied to numerous (i.e.,
hundreds) of correlated input features in \citet{Hastie:1995p2589}.
Here we extend the results of the latter work to include sparse structured
regression methods that can be fit efficiently to hundreds of thousands
of input features. }

\textcolor{black}{More formally, OS finds an optimal scoring function
$\theta:g\rightarrow\mathbb{R}$ that maps classes $g\in\{1,...J\}$
into the real numbers. In the case of a multi-class classification
using the Elastic Net, we can apply OS to yield estimates}

\textcolor{black}{
\begin{eqnarray}
(\widehat{\Theta},\widehat{\beta}) & = & \kappa\ \underset{\Theta,\beta}{\text{argmin}}\ \|Y\Theta-X\beta\|_{2}^{2}+\lambda_{1}\|\beta\|_{1}+\lambda_{2}\|\beta\|_{2}^{2}\label{eq:5}\\
 &  & \text{subject to }n^{-1}\|Y\Theta\|_{2}^{2}=1,
\end{eqnarray}
where $\Theta$ is a matrix that yields the optimal scores when applied
to indicator matrix $Y$, and where we add the constraint \eqref{eq:5}
to avoid degenerate solutions \citep{Grosenick:2008p2789}. Given
that this is just a sparse version of PDA \citep{Hastie:1995p2589},
we have called this combination Sparse Penalized Discriminant Analysis
(SPDA). It has also recently been called Sparse Discriminant Analysis
(SDA) \citet{Clemmensen2011} (and for an interesting alternative
approach for constructing sparse linear discriminant classifiers,
see \citet{Witten2011}).}

\textcolor{black}{For simplicity, we consider only a local spatiotemporal
smoothing penalty in the current study, although using more elaborate
spatial/temporal coordinates would follow similar logic. The SPDA-GraphNet
is defined as
\begin{eqnarray}
(\widehat{\Theta},\widehat{\beta}) & = & \kappa\ \underset{\Theta,\beta}{\text{argmin}}\ \|Y\Theta-X\beta\|_{2}^{2}+\lambda_{1}\|\beta\|_{1}+\lambda_{G}\|\beta\|_{G}^{2}\label{eq:10}\\
 &  & \text{subject to \ensuremath{n^{-1}\|}Y\ensuremath{\Theta\|_{2}^{2}}=1.}
\end{eqnarray}
It is important to note that the direct equivalence between penalized
OS and penalized LDA has only recently been proven in the binary classification
case, and does not hold for multi-class classification problems \citep{Merchante2012ICML}.
However, both approximate methods that iteratively minimize over $\Theta$
and $\beta$ \citep{Clemmensen2011} and equivalent methods based
on the Group Lasso \citep{Merchante2012ICML} could be used with GraphNet
regression methods to build multi-class GraphNet classifiers. We note
that in the binary classification case there are at least two options
to turn regression methods into classifiers: Optimal Scoring and logistic
regression (see e.g. \citealt{Friedman2010}). In the case of multiple
classes, the approaches of \citep{Clemmensen2011,Merchante2012ICML}
provide LDA or LDA-like classifiers. Sparse multinomial regression
could also be used in the multi-class case. Any of these approaches
may be used to turn GraphNet regression methods into GraphNet classifiers.
Because Optimal Scoring converts regression methods into equivalent
linear discriminant classifiers, it allows us to combine notions from
regression such as degrees of freedom with notions from discriminant
analysis such as class visualization in the discriminant space using
discriminant coordinates and trial-by-trial posterior probabilities
for individual observations \citep{Hastie:1995p2589}. This, and its
greater computational simplicity over logistic and multinomial regressions,
make OS an appealing approach.}
\end{singlespace}

\textcolor{black}{\medskip{}
}

\subsubsection{\textcolor{black}{Turning regression methods into classifiers: relating
Support Vector Machines (SVM) to penalized regression}}

\textcolor{black}{\medskip{}
}

In addition to the LDA and logistic/multinomial approaches to classification,
maximum margin classifiers like SVM have been very successful. As
we will also be developing a Support Vector GraphNet (SVGN) variant
below, we briefly discuss how support vector machines can be related
to regression methods like those described above. If the data is centered
such that an intercept term can be ignored, the SVM solution can be
written 
\[
\widehat{\beta}=\underset{\beta}{\text{argmin}}\ \sum_{i=1}^{n}(1-y_{i}x_{i}^{T}\beta)_{+}+(\lambda/2)\|\beta\|_{2}^{2},
\]
where $(\cdot)_{+}$ indicates taking the positive part of the quantity
in parentheses. In this function estimation formulation of the SVM
problem, we see the similarity to the penalized regression methods
above: the only difference is that the usual squared error loss $L(y_{i},x_{i},\beta)=(y_{i}-x_{i}^{T}\beta)^{2}$
has been replaced by the {}``hinge loss'' function $L_{H}(y_{i},x_{i},\beta)=(1-y_{i}x_{i}^{T}\beta)_{+}$
. This function is non-differentiable, and more recent work \citep{WangZou2008}
uses a differentiable {}``Huberized hinge loss'' (Figure 2a), which
we will discuss in greater detail below. The important point here
is that formulating the SVM problem as a loss term and a penalty term
reveals how we might build an SVM with more general penalization,
such as that used in GraphNet regression methods above.\textcolor{black}{\medskip{}
}

\begin{singlespace}

\subsection{Novel extensions of GraphNet methods\textcolor{black}{\medskip{}
}}
\end{singlespace}

\subsubsection{\textcolor{black}{Robust GraphNet and Adaptive Robust GraphNet\medskip{}
}}

\begin{singlespace}
\textcolor{black}{More generally, we can formulate the penalized regression
problem of interest as minimizing the penalized empirical risk }\textbf{\textcolor{black}{$\mathcal{R}_{p}(\beta)$}}\textcolor{black}{{}
as a function of the coefficients, so that 
\begin{equation}
\widehat{\beta}=\underset{\beta}{\text{argmin}\ }\mathcal{R}_{p}(\beta)=\underset{\beta}{\text{argmin}}\ \mathcal{R}(y,\widehat{y})+\lambda\mathcal{P}(\beta),\label{eq:6}
\end{equation}
where $\widehat{y}$ is the estimate of response variable $y$ (note
$\widehat{y}=X\widehat{\beta}$ in the linear models we consider)
and $\mathcal{R}(y,\widehat{y})=n^{-1}\sum_{i=1}^{n}L(y_{i},\widehat{y}_{i})$
is the average of the loss function over the training data (the {}``empirical
risk'') of the loss function $L(y_{i},\widehat{y}_{i})$ that penalizes
differences between the estimated and true values of $y$ at the $i$th
observation. For example, in  \eqref{eq:3}--\eqref{eq:5} we used
$\mathcal{R}(y,\widehat{y})=\|y-\widehat{y}\|_{2}^{2}=\sum_{i=1}^{n}(y_{i}-\widehat{y}_{i})^{2}$
({}``squared error loss''). While squared error loss enjoys many
desirable properties under the assumption of Gaussian noise, it is
sensitive to the presence of outliers.}

\textcolor{black}{Outlying data points are an important consideration
when modeling fMRI data, in which a variety of factors ranging from
residual motion artifacts to field inhomogeneities can cause some
observations to fall far from the sample mean. In the case of standard
squared-error loss (as in equations \eqref{eq:2}--\eqref{eq:5}),
these outliers can have undue influence on the model fit due to the
quadratically increasing penalty on the residuals (see Figure 2a).
A standard solution in such cases is to use a robust loss function,
such as the Huber loss function \citep{Huber2009},
\begin{eqnarray}
\mathcal{R}_{H}(y,\widehat{y};\delta) & =n^{-1} & \sum_{i=1}^{n}L_{\delta}(y_{i}-\widehat{y}_{i})\label{eq:7}\\
\text{where } &  & L_{\delta}(y_{i}-\widehat{y}_{i})=\begin{cases}
(y_{i}-\widehat{y}_{i})^{2}/2 & \text{for }|y_{i}-\widehat{y}_{i}|\leq\delta\\
\delta|y_{i}-\widehat{y}_{i}|-\delta^{2}/2 & \text{for }|y_{i}-\widehat{y}_{i}|>\delta
\end{cases}.\nonumber 
\end{eqnarray}
This function penalizes residuals quadratically when they are less
than or equal to parameter $\delta$, and linearly when they are larger
than $\delta$ (Figure 2a). A well specified $\delta$ can thus significantly
reduce the effects of large residuals (outliers) on the model fit,
as they no longer have the leverage resulting from a quadratic penalty.
As $\delta\rightarrow\infty$ (or practically, when it becomes larger
than the most outlying residual) we recover the standard squared-error
loss. }

\textcolor{black}{Since GraphNet uses squared-error loss, it can now
be modified to include a robust penalty like the Huber loss defined
above. Replacing the squared error loss function with the loss function
\eqref{eq:7} yields }

\textcolor{black}{
\begin{equation}
\widehat{\beta}=\kappa\ \underset{\beta}{\text{argmin}\ }\mathcal{R}_{H}(y,X\beta;\delta)+\lambda_{1}\|\beta\|_{1}+\lambda_{G}\|\beta\|_{G}^{2}.\label{eq:11}
\end{equation}
}The Adaptive Robust GraphNet is then a straightforward generalization
(following section 2.1.5; see also the next section) 
\end{singlespace}

\textcolor{black}{The SPDA-RGN classifier can be defined like the
standard GraphNet classifier \eqref{eq:10}. However, the SPDA-RGN
classifier now has an additional hyperparameter to be estimated (or
assumed). Specifically, the value of $\delta$ determines where the
loss function switches from quadratic to linear (Figure 2a). Further,
the loss function on the residuals is no longer quadratic and therefore
could slow down optimization convergence . We discuss a solution to
this issue next.\medskip{}
}

\begin{singlespace}

\subsubsection{\textcolor{black}{Infimal convolution for non-quadratic loss functions}}
\end{singlespace}

\textcolor{black}{\medskip{}
}

In order to solve both the Robust GraphNet, Adaptive Robust GraphNet,
and Support Vector GraphNet problems efficiently, we introduce a general
method for solving coordinate-wise descent problems with smooth, non-quadratic
convex loss functions as penalized least squares problems in an augmented
set of variables.

\textcolor{black}{Convergence speed of subgradient methods such as
coordinate-wise descent can be substantially improved when the loss
function takes a quadratic form, while non-quadratic loss functions
can take numerous iterations to converge for each coefficient, significantly
increasing computation time. However, we can circumvent these problems
and extend the applicability of coordinate-wise descent methods using
a trick from convex analysis to rewrite these loss functions as quadratic
forms in an augmented set of variables. This method is called infimal
convolution \citep{Rockafellar1970}, and is defined as}

\begin{flushleft}
\textcolor{black}{
\begin{equation}
(f\star_{\text{inf}}g)(x):=\inf_{y}\{f(x-y)+g(y)|y\in\mathbb{R}^{n}\},\label{eq:infimal convolution}
\end{equation}
where $f$ and $g$ are two functions of $x\in\mathbb{R}^{p}$. In
this way it is possible to rewrite the $i$th term in the the Huber
loss function \eqref{eq:7} as the infimal convolution of the squared
and absolute-value functions applied to the $i$th residual $r_{i}$:
\begin{equation}
\rho_{\delta}(r_{i})=((1/2)(\cdot)^{2}\star_{\text{inf}}|\cdot|)(r_{i})=\inf_{a_{i}+b_{i}=r_{i}}a_{i}^{2}/2+\delta|b_{i}|,\label{eq:infimal conv for huber loss}
\end{equation}
where $r_{i}=y_{i}-(X\widehat{\beta})_{i}$ (note that a dot $(\cdot)$
is used to indicate the functional nature of the expression without
having to add additional dummy variables). This yields the augmented
estimation problem 
\begin{equation}
(\widehat{\alpha},\widehat{\beta})=\underset{\alpha,\beta}{\text{argmin}}\ (1/2)\|y-X\beta-\alpha\|_{2}^{2}+\lambda_{G}\beta^{T}G\beta+\delta\|\alpha\|_{1}+\lambda_{1}\|\beta\|_{1},\label{eq:infimal huberized graphnet}
\end{equation}
where we have introduced the auxiliary variables $\alpha\in\mathbb{R}^{n}$.
Considering the residuals $r_{i}$, the first term in the objective
of \eqref{eq:infimal huberized graphnet} can be written $(1/2)\|y-X\beta-\alpha\|_{2}^{2}=(1/2)\sum_{i}(r_{i}-\alpha_{i})^{2},$
and thus each $\alpha_{i}$ can directly reduce the residual sum of
squares corresponding to a single observation by taking a value close
to $r_{i}$. Since for some $\delta$ the penalty $\delta\|\alpha\|_{1}$
requires the $\alpha$ vector to be $k$-sparse, this formulation
intuitively allows a linear rather that quadratic penalty to be placed
on $k$ of the residuals (with $k$ tuned by choice of $\delta,$
as expressed in the Huber loss formulation). These will correspond
to those observations with the most leverage (the most {}``outlying''
points). We can then rewrite \eqref{eq:infimal huberized graphnet}
as 
\begin{eqnarray}
\widehat{\gamma} & = & \underset{\gamma}{\text{argmin}}\ (1/2)\|y-Z\gamma\|_{2}^{2}+\lambda_{G}\gamma^{T}G'\gamma+\sum_{j=1}^{p+n}w_{j}|\gamma_{j}|\label{eq:infimal huberized graphnet rewritten}\\
\text{} &  & Z=[X\ \ I_{n\times n}],\ \ \gamma=[\beta\ \ \alpha],\ \ w_{j}=\begin{cases}
\lambda_{1} & j=1,..,p\\
\delta & j=p+1,\ldots,p+n,
\end{cases}\nonumber \\
 &  & G'=\left[\begin{array}{ll}
G & 0_{1\times n}\\
0_{n\times1} & 0_{n\times n}
\end{array}\right]\in S_{+}^{(p+n)\times(p+n)},\nonumber 
\end{eqnarray}
where $S_{+}^{m\times m}$ is the set of positive semidefinite $m\times m$
matrices. This is just a GraphNet problem in an augmented set of $p+n$
variables, and so can be solved using the fast coordinate-wise descent
methods discussed in section 2.4 below. After solving for augmented
coefficients $\widehat{\gamma}$ we can simply discard the last $n$
coefficients to yield $\widehat{\beta}$. A similar approach can be
taken with the hinge-loss of a support vector machine classifier (as
we show next), or more generally with any loss function decomposable
into an infimal convolution of convex functions (see Appendix). The
Adaptive Robust GraphNet is easily obtained by letting 
\[
w_{j}=\begin{cases}
\lambda_{1}^{*}\widehat{w}_{j} & j=1,..,p\\
\delta & j=p+1,\ldots,p+n
\end{cases}
\]
in \eqref{eq:infimal huberized graphnet rewritten} (see section 2.1.5
for more details on adaptive estimation).\medskip{}
}
\par\end{flushleft}

\begin{singlespace}

\subsubsection{\textcolor{black}{Huberized Support Vector Machine (SVM) GraphNet
for classifications}}
\end{singlespace}

\textcolor{black}{\medskip{}
}

\textcolor{black}{In the $p\gg n$ classification problem, maximum-margin
classifiers such as the support vector machine (SVM) often perform
exceedingly well in terms of classification accuracy, but do not yield
readily interpretable coefficients. For this reason we also develop
a sparse SVM with graph constraints, the Support Vector GraphNet (SVGN),
related to the {}``Hybrid Huberized SVM'' of \citet{WangZou2008}
as an alternative to the SPDA method. Using a {}``Huberized-hinge''
loss function $\mathcal{R}_{HH}$ (see below) on the fit residuals,
we have }

\begin{singlespace}
\textcolor{black}{
\begin{equation}
\widehat{\beta}=\kappa\ \underset{\beta}{\text{argmin}\ }\mathcal{R}_{HH}(y^{T}X\beta;\delta)+\lambda_{1}\|\beta\|_{1}+\lambda_{G}\|\beta\|_{G}^{2},\label{eq:12-1}
\end{equation}
where $y\in\{-1,1\}$, and letting $\widehat{y}=X\widehat{\beta}$
be the estimates of the target variable,}
\end{singlespace}

\textcolor{black}{
\begin{eqnarray}
\mathcal{R}_{HH}\left(y,\widehat{y};\delta\right) & = & n^{-1}\sum_{i=1}^{n}L_{\delta}(y_{i},\widehat{y}_{i})\label{eq:huber loss}\\
\text{where} &  & L_{\delta}(y_{i},\widehat{y}_{i})=\begin{cases}
\left(1-y_{i}\widehat{y}_{i}\right)^{2}/2\delta & \text{for }1-\delta<y_{i}\widehat{y}_{i}\leq1\\
1-y_{i}\widehat{y}_{i}-\delta/2 & \text{for }y_{i}\widehat{y}_{i}\leq1-\delta\\
0 & \text{for }y_{i}\widehat{y}_{i}>1,
\end{cases}\nonumber 
\end{eqnarray}
which is the Huberized-hinge loss of \citet{WangZou2008}. As with
the Huber loss, there is an additional hyperparameter $\delta$ to
be estimated or assumed. In this case, $\delta$ determines where
the hinge-loss function switches from the quadratic to the linear
regime (see Figure 2a). This problem's loss function can also be written
using infimal convolution to yield a more convenient quadratic objective
term (see Appendix). Finally, we discuss a heuristic alternative to
adaptive methods for adjusting nonzero coefficient magnitudes to match
the scale of the original data. This approach can be used with any
of the above methods. \medskip{}
}

\begin{singlespace}

\subsubsection{\textcolor{black}{Effective degrees of freedom for GraphNet estimators}}
\end{singlespace}

\textcolor{black}{\medskip{}
}

Following results for the Lasso \citet{ZouHastieDFs} and the Elastic
Net \citet{VanDerKooij2007}, the effective degrees of freedom $\widehat{df}$
for the GraphNet regression are given by the trace of the {}``hat
matrix'' $H_{\lambda_{G}}(\mathcal{A})$ for the GraphNet estimator:
\[
\widehat{df}=\text{tr}(H_{\lambda_{G}}(\mathcal{A}))=\text{tr}\left(X_{\mathcal{A}}\left(X_{\mathcal{A}}^{T}X_{\mathcal{A}}+\lambda_{G}G\right)^{-1}X_{\mathcal{A}}^{T}\right),
\]
where $X_{\mathcal{A}}$ denotes the columns of $X$ containing just
the {}``active set'' (those variables with nonzero coefficients
corresponding to a particular choice of $\lambda_{1}$). This quantity
is very useful in calculating standard model selection criteria such
as the Akaike Information Criterion (AIC), Bayesian Information Criterion
(BIC), Mallow's $C_{p}$, Generalized Cross Validation (GCV), and
others. Importantly, it can also be used for the various GraphNet
methods, as each of these is solved as an equivalent GraphNet problem
(for example, equation \ref{eq:infimal huberized graphnet rewritten})
for the Adaptive Robust GraphNet. \textcolor{black}{\medskip{}
}

\begin{singlespace}

\subsubsection{\textcolor{black}{Rescaling coefficients to account for {}``double
shrinking''}}
\end{singlespace}

\textcolor{black}{\medskip{}
}

\begin{singlespace}
\textcolor{black}{The Elastic Net was originally formulated by \citet{ZouHastie}
in both {}``naive'' and rescaled forms. The authors noted that a
combination of $\ell_{1}$ and $\ell_{2}$ penalties can {}``double
shrink'' the coefficients. To correct this they proposed rescaling
the {}``naive'' solution by a factor of $\kappa=1+\lambda_{2}$
\citep{ZouHastie}. Heuristically, the aim is to retain the desirable
variable selection properties of the Elastic Net while rescaling the
coefficients to be closer to the original scale. However, as this
result is derived for an orthogonal design, it is not clear that $\kappa=1+\lambda_{2}$
is the correct multiplicative factor if the data are collinear, and
this can complicate the problem of choosing a final set of coefficients.
Following the arguments of \citet{ZouHastie}, for GraphNet regression
we might rescale each coefficient by $\kappa_{j}=k(\widehat{\Sigma}_{jj}+\lambda_{G}G_{jj})$
where $k$ is the number of iterations used in the coordinate-wise
descent optimization (and thus the number of times shrinkage related
to $G$ is applied, see equation \ref{eq:15-2} and derivations in
Appendix) and where $\widehat{\Sigma}=X^{T}X$. In the case of an
orthogonal design and $G=I$ we would have $\widehat{\Sigma}=1$ and
thus $\kappa_{j}=1+\lambda_{G}$---reducing to the Elastic Net rescaling
employed in \citet{ZouHastie}.}

\textcolor{black}{A simpler alternative is to fit the Elastic Net,
generating a fitted response $\hat{y}$, and then to regress $y$
on $\hat{y}$. In particular, solving the simple linear regression
problem
\[
y=\kappa\widehat{y}=\kappa X\widehat{\beta},\ \kappa\in\mathbb{R}
\]
yields an estimate $\widehat{\kappa}$ that can be used to rescale
the coefficients obtained from fitting the Elastic Net (Daniela Witten
and Robert Tibshirani, personal communication). The intuitive motivation
for this heuristic is that it will produce a $\widehat{\kappa}$ that
puts $\widehat{\beta}$ and $\widehat{y}$ on a reasonable scale for
fitting $y$.}

\textcolor{black}{Besides its simplicity, the principal advantage
of this approach is that it requires no analytical knowledge about
the amount of shrinkage that occurs as $\lambda_{G}$ is increased.
This is particularly appealing because the same strategy of regressing
$\hat{y}$ on $y$ can be used with more general problems with more
complicated forms, such as the Adaptive Robust GraphNet, where the
additional shrinkage caused by the graph penalty can be corrected
in this way.}
\end{singlespace}

Finally, we note that over-shrinking is not necessarily bad for classification
accuracy. Indeed it may improve accuracy due to the rather complicated
relationship between bias and variance in the of classification (for
an excellent discussion in the context of 0-1 loss see \citet{Friedman1997}).
The focus on recovering good estimates of coefficent magnitude in
this section is thus most relevant to regression and to situations
in which correct estimates of coefficient magnitude are important.

\textcolor{black}{\medskip{}
}

\begin{singlespace}

\subsection{\textcolor{black}{Interpreting GraphNet regression and classification
\medskip{}
}}
\end{singlespace}

\subsubsection{\textcolor{black}{Interpreting GraphNet parameters: dual variables
as prices\medskip{}
}}

\textcolor{black}{The GraphNet problem expressed in equation \eqref{eq:8}
derives from a constrained maximum likelihood problem, in which we
want to maximize the likelihood of the parameters given the data,
subject to some hard constraints on the solution---specifically, that
they are sparse and structured (in the sense that their $\ell_{1}$
and graph-weighted $\ell_{2}$ norms are less than or equal to some
constraint size). For concave likelihoods (as in generalized linear
models and the cases considered above), this is a constrained convex
optimization problem 
\begin{eqnarray}
\underset{\beta}{\text{maximize}} &  & \text{loglik}(\beta|X,y)\label{eq: constrained mle objective}\\
\text{subject to} &  & \|\beta\|_{1}\leq c_{1}\label{eq: constrained mle l1}\\
 &  & \|\beta\|_{G}^{2}\leq c_{G},\label{eq: constrained mle graphpen}
\end{eqnarray}
where $c_{1}\in\mathbb{R}_{+}$ and $c_{G}\in\mathbb{R}_{+}$ set
hard bounds on the size of the coefficients in the $\ell_{1}$ and
$\ell_{G}$ norms, respectively. A standard approach for solving such
problems is to relax the hard constraints to linear penalties \citep{BoydVandenberghe2004}
and consider just those terms containing $\beta$, giving the {}``Lagrangian''
form of the GraphNet problem}

\textcolor{black}{
\begin{equation}
\widehat{\beta}\underset{\beta}{=\text{argmin}}\ -\text{loglik}(\beta|X,y)+\lambda_{1}\|\beta\|_{1}+\lambda_{G}\|\beta\|_{G}^{2},\ \lambda_{1},\lambda_{G}\in\mathbb{R}_{+},\label{eq:penalized loglikelihood}
\end{equation}
which contains a negative likelihood term that measures misfit to
the data as well as the two penalties characteristic of GraphNet estimators. }

\textcolor{black}{In this Lagrangian formulation, the dual variables
$\lambda_{1}$ and $\lambda_{G}$ represent (linear) costs in response
to a violation of the constraints. Since we solve problem \eqref{eq:penalized loglikelihood},
$c_{1}$ and $c_{G}$ are effectively zero, and we are penalized for
any deviation of the coefficients from zero. This leads to one interpretation
of $\lambda_{1}$ and $\lambda_{G}$: they are prices that we are
willing to pay to improve the likelihood at the expense of a less
sparse or less structured solution, respectively. For this reason,
examining fit sensitivity to different values of $\lambda_{1}$ and
$\lambda_{G}$ tells us about underlying structure in the data. For
example, if the task-related neural activity was very sparse and highly
localized in a few uncorrelated voxels, then we should be willing
to pay more for sparsity and less for smoothness (i.e., large $\lambda_{1}$,
small $\lambda_{G}$). In contrast, if large smooth and correlated
regions underlie the task, then tolerating a large $\lambda_{G}$
could substantially improve the fit. To explore such possibilities,
we can plot test rates from cross validations at different combinations
of  parameters. Figure 6 shows plots of median test classification
rates as a function of $\lambda_{1}$ and $\lambda_{G}$ over the
parameter grid on which the various GraphNet classifiers were fit.
We see that there are regions in the $(\lambda_{1},\lambda_{G})$
parameter space that clearly result in better median classification
test rates, corresponding to fits with particular levels of smoothness
and sparsity.}

\textcolor{black}{\medskip{}
}

\subsubsection{\textcolor{black}{Interpreting GraphNet coefficients }}

\textcolor{black}{\medskip{}
}

Problem \eqref{eq:penalized loglikelihood} can also be arrived at
from a Bayesian perspective as a maximum a posteriori (MAP) estimator.
In this case, the form of the penalty \textbf{$\mathcal{P}(\beta)$
}is related to one's prior beliefs about the structure of the  coefficients.
\textcolor{black}{For example, under the well-known equivalence of
penalized regression techniques and posterior modes, the Elastic Net
penalty corresponds to the prior 
\[
p_{\lambda_{1},\lambda_{2}}(\beta)\propto\exp\left\{ -\left(\lambda_{1}\|\beta\|_{1}+\lambda_{2}\|\beta\|_{2}^{2}\right)\right\} 
\]
\citep{ZouHastie}. The GraphNet penalty thus corresponds to the prior
distribution
\begin{eqnarray}
p_{\lambda_{1},\lambda_{G}}(\beta) & \propto & \exp\left\{ -\left(\lambda_{1}||\beta||_{1}+\lambda_{G}\beta^{T}G\beta\right)\right\} \nonumber \\
 & \propto & \prod_{i=1}^{p}\exp\left\{ -\lambda_{1}|\beta_{j}|\right\} \prod_{i=1}^{p}\exp\left\{ -\lambda_{G}\sum_{i\sim j}\beta_{i}G_{ij}\beta_{j}\right\} ,\label{eq: bayesian interp}
\end{eqnarray}
where $i\sim j$ denotes that node $i$ in the graph $G$ is adjacent
to node $j$. Therefore, the GraphNet problems are also equivalent
to a MAP estimator of the coefficients with a prior consisting of
a convex combination of a global Laplacian (double-exponential) and
a local Markov Random Field (MRF) prior. In other words, GraphNet
methods explicitly take into account prior beliefs about coefficients
being globally sparse but locally structured. }

\begin{singlespace}

\subsection{\textcolor{black}{Optimization and computational considerations\medskip{}
}}
\end{singlespace}

\begin{singlespace}

\subsubsection{\textcolor{black}{Coordinate-wise descent and active set methods}}
\end{singlespace}

\textcolor{black}{\medskip{}
}

\begin{singlespace}
\textcolor{black}{Fitting regression methods to whole-brain fMRI data
requires efficient computational methods, particularly when they must
be cross-validated over a grid of possible parameter values. For instance,
in the shopping example described in greater detail below (section
2.5), 26,630 input features (voxels) at each of 7 time points are
used to classify future choices to purchase a product or not. Fitting
the Adaptive Robust GraphNet using leave-one-subject-out (LOSO) cross-validation
(i.e., 25 fits) for each realization of possible parameter values
over this $90\times5\times6\times10\times3$ grid of possible parameters
$\{\lambda_{1},G,\lambda_{G},\delta,\lambda_{1}^{*}\}$ requires $2,025,000$
model fits on $1,882$ observations of $186,410$ input features. }

\textcolor{black}{To efficiently fit GraphNet methods with millions
of parameter combinations over hundreds of thousands of input features,
we formulated the minimization problem (i.e., equations \eqref{eq:8},
\eqref{eq:11}, and \eqref{eq:12-1}) as a coordinate-wise optimization
procedure \citep{Tseng1988,Tseng2001} using active set methods \citep{Friedman2010}.
This approach fit one coefficient value at a time ({}``coordinate-wise''
descent), holding the rest constant, and kept an {}``active set''
of nonzero coefficients. Fitting was initiated with a large value
of $\lambda_{1}$ (corresponding to all coefficients being zero),
and then slowly decreased $\lambda_{1}$ to allow more and more coefficients
into the model fit. This procedure thus considered an {}``active
set'' of the model coefficients at each coordinate-wise update, rather
than all 186,410 inputs. Occasional sweeps though all the coefficients
were made to search for new variables to include, as in \citet{Friedman2010}.
Model fitting terminated before $\lambda_{1}$ reached zero, since
fitting a fully dense set of coefficients is computationally expensive
and known to produce poor estimates \citep{Hastie:2009p2681,Friedman2010}.
Various heuristics and model selection criteria may be used for choosing
a stopping point, for example, stopping once the AIC or BIC for the
model stops decreasing and starts increasing. AIC is known to be over-inclusive
in model selection, and is therefore a more conservative stopping
point.}

\textcolor{black}{Coordinate-wise descent is guaranteed to converge
for GraphNet methods because they are all of the form 
\begin{equation}
\underset{\beta}{\text{argmin}}\ f(\beta_{1},...,\beta_{p})=\underset{\beta}{\text{argmin}}\ g(\beta_{1},..,\beta_{p})+\sum_{j=1}^{p}h(\beta_{j}),\label{eq:14}
\end{equation}
where $g(\beta_{1},..,\beta_{p})$ is a convex, differentiable function
(e.g., squared-error and Huber loss plus the quadratic penalty $||\beta||_{G}^{2}$),
and where each $h(\beta_{j})$ is a convex (but not necessarily differentiable)
function (e.g., the $\ell_{1}$ penalty). If the convex, non-differentiable
part of the penalty function is separable in coordinates $\beta_{j}$
(as is true of $||\beta||_{1}=\sum_{j=1}^{p}|\beta_{j}|$), then coordinate
descent converges to a global solution of the minimization problem
\citep{Tseng2001}. In the case of Huber loss or Huberized-hinge loss,
the two-part loss function can be written as a single quadratic loss
function using infimal convolution as described in section 2.2.3.
For instance, consider the coordinate-wise updates for the standard
GraphNet problem given in equation \eqref{eq:8}. Letting $\hat{y}=\tilde{X}\tilde{\beta}+X._{j}\beta_{j}$
(where $\tilde{X}=X._{\neq j}$ is the matrix $X$ with the $j$th
column removed, and $\tilde{\beta}=\beta_{\neq j}$ the coefficient
vector with the $j$th coefficient removed), the subdifferential of
the risk with respect to only the $j$th coefficient (holding the
others fixed) is 
\begin{equation}
\partial_{\beta_{j}}\mathcal{R}_{p}=-X._{j}^{T}y+X._{j}^{T}\tilde{X}\tilde{\beta}+X._{j}^{T}X._{j}\beta_{j}+(\lambda_{2}/2)\tilde{\beta}^{T}(G_{\neq j}.)._{j}+\lambda_{2}G_{jj}\beta_{j}+(\lambda_{1}/2)\text{\ensuremath{\Gamma}}(\beta_{j}),\label{eq:15-1}
\end{equation}
where the set-valued function $\Gamma(\beta_{j})=-1$ if $\beta_{j}<0$,
$\Gamma(\beta_{j})=1$ if $\beta_{j}>0$ and $\Gamma(\beta_{j})\in[-1,1]$
if $\beta_{j}=0$. If we let $\Gamma(\beta_{j})=\text{sign}(\beta_{j}),$
in equation \eqref{eq:15-1} (which is always a particular subgradient
in the subdifferential of the risk), then the coordinate update iteration
for the $j$th coefficient estimate is 
\begin{equation}
\hat{\beta}_{j}\leftarrow\frac{S\left(X._{j}^{T}(y-\tilde{X}\tilde{\beta})-(\lambda_{2}/2)\tilde{\beta}^{T}(G_{\neq j}.)._{j},\ \lambda_{1}/2\right)}{X._{j}^{T}X._{j}+\lambda_{2}G_{jj}},\label{eq:15-2}
\end{equation}
where 
\begin{equation}
S(x,\gamma)=\text{sign}(x)(|x|-\gamma)_{+}\label{eq:soft-thresh}
\end{equation}
is the soft-thresholding function \citep{DonohoUST,Friedman:2007p36}.
Note that if graph $G=I$, and the columns of $X$ are standardized
to have unit norm, then the coordinate-wise Elastic Net update is
recovered \citep{VanDerKooij2007,FriedmanPWCO}.}
\end{singlespace}

\textcolor{black}{\medskip{}
}

\begin{singlespace}

\subsubsection{\textcolor{black}{Computational complexity}}
\end{singlespace}

\textcolor{black}{\medskip{}
}

\textcolor{black}{A closer look at equation \eqref{eq:15-2} reveals
that if the variables are standardized (such that $X._{j}^{T}X._{j}=1$)
then the $(c+1)$st coefficient update for the $j$th coordinate can
be rewritten 
\begin{equation}
\hat{\beta}_{j}^{(c+1)}\leftarrow S\left(\sum_{i=1}^{N}x_{ij}r_{i}^{(c)}+\hat{\beta}_{j}^{(c)}-(\lambda_{2}/2)\sum_{k\neq j}\beta_{k}G_{kj},\ \lambda_{1}/2\right)/(1+\lambda_{2}G_{jj}),\label{15-3}
\end{equation}
where $r=y-\hat{y}$ is the vector of residuals. Letting $m$ be the
number of off-diagonal nonzero entries in $G$ and initializing with
$\hat{\beta}_{j}^{(0)}=0$ for all $j$ and $r^{(0)}=y$, the first
sweep through all $p$ coefficients will take $O(pn)+O(m)$ operations.
Once $a_{1}$ variables are included in the active set, $q$ iterations
are performed according to \eqref{15-3} until the new estimates converge,
at which point $\lambda_{1}$ is decreased incrementally and another
$O(pn)$ sweep is made through the coefficients to find the next active
set with $a_{2}$ variables (using the previous estimate as a warm
start to keep $q$ small). This procedure is repeated for $l$ values
of $\lambda_{1}$, until the fit stops improving or a pre-specified
coefficient density is reached. Let $a=\sum_{i=1}^{l}a_{i}$ denote
the total number of coefficients updates over all $l$ fits. The total
computational complexity is then $O(lpn)+O(lm)+O(aq)$. Thus if $G$
is relatively sparse (so $m$ is small) and if it requires few iterations
for coefficients in an active set to converge ($q$ small)---which
is true if the unpenalized loss function is quadratic---then the computational
complexity is dominated by the $O(lpn)$ term representing the sweep
through the coefficients necessary to find the next active set for
each new value of $\lambda_{1}$. We note that this suggests that
including a screening procedure such as the STRONG rules \citep{Tibs2012}
could further speed up fitting in this context. Either making $G$
dense or decreasing $\lambda_{1}$ until $a$ becomes large can cause
the other complexity terms to play a significant role and slow the
speed of the algorithm. For example, if $G$ is dense, then $m=p^{2}-p$
and the $O(lm)$ term will dominate.}

\textcolor{black}{\medskip{}
}

\begin{singlespace}

\subsubsection{\textcolor{black}{Cross validation, classification accuracy, and
parameter tuning}}
\end{singlespace}

\textcolor{black}{\medskip{}
}

For training and test data, trials for each subject were resampled
within-subject to consist of 80 trials with exactly 40 purchases.
If the subject originally had more than 40 purchases, sampling without
replacement was used to select 40. If the subject originally had fewer
than 40 purchases, sampling with replacement was used to select 40.
Similar sampling was used to select exactly 40 trials without purchases.
This resampling scheme ensured that the trials for each subject were
balanced between purchasing and not purchasing. Further, because our
cross-validation schemes defined folds on the subject level, this
ensured that every training and test set in the cross-validation was
also balanced. 

\begin{singlespace}
\textcolor{black}{For the cross-validation, the range for these grid
values was chosen based off of a few preliminary fits. This grid is
very large, and with the refitting involved in cross-validation, resulted
in millions of fits. The smoothness of the rates as a function of
the parameters (see Figure 6) suggests that smaller grids are likely
better suited to most applications, and we anticipate that more efficient
adaptive approaches to parameter search---such as focused grid search
methods \citep{Jimenez2009} or sampling methods inspired by Bayesian
approaches to similar problems---will ultimately prove superior. We
leave these refinements to future work. The grid values used here
are given in the Appendix.}
\end{singlespace}

In order to choose a final set of coefficient estimates from multiple
fits across cross-validation folds, we took the element-wise median
of the coefficient vectors across the folds. Thus a feature corresponding
to a particular voxel at a particular TR would have to appear (be
nonzero) in more than half of the 25 cross-validation folds in order
to be included in the final coefficient estimate used in the out-of-sample
(OOS) analysis. There are several justifications for taking the median
across folds: (1) the median preserves sparsity, (2) the median is
the appropriate maximum likelihood estimator for the double-exponential
(Laplacian) distribution that corresponds to the $\ell_{1}$ sparsity
prior on the coefficients (see discussion in \citealt{Grosenick:2008p2789}),
(3) such a procedure is closely related to the Median Probability
Model, which is the model consisting of those variables that have
posterior probability $\geq0.5$ of being in a model, and which has
been shown to have optimal predictive performance for linear models
\citep{BarbieriBerger2004}, and (4) it is similar to other recently-developed
model selection procedures for sparse models such as Stability Selection
\citealt{MeinshausenBuhlmann2010} that use the number of times a
variable appears across multiple sparse fits to resampled data in
order to significantly improve model selection. Further, we have found
this approach to be quite effective in practice (see the out-of-sample
results that follow). Note that such inclusion of a variable only
if it appears in more than half of the 25 cross-validation folds is
a natural means of imposing some {}``reliability'' or {}``stability''
on the coefficients. 

\textcolor{black}{\medskip{}
}

\subsection{\textcolor{black}{Application: Predicting buying behavior using fMRI\medskip{}
}}

\begin{singlespace}

\subsubsection{\textcolor{black}{Subjects and SHOP task}}
\end{singlespace}

\textcolor{black}{\medskip{}
}

\begin{singlespace}
\textcolor{black}{}
\begin{figure}[t]
\begin{centering}
\textcolor{black}{\includegraphics[scale=0.8]{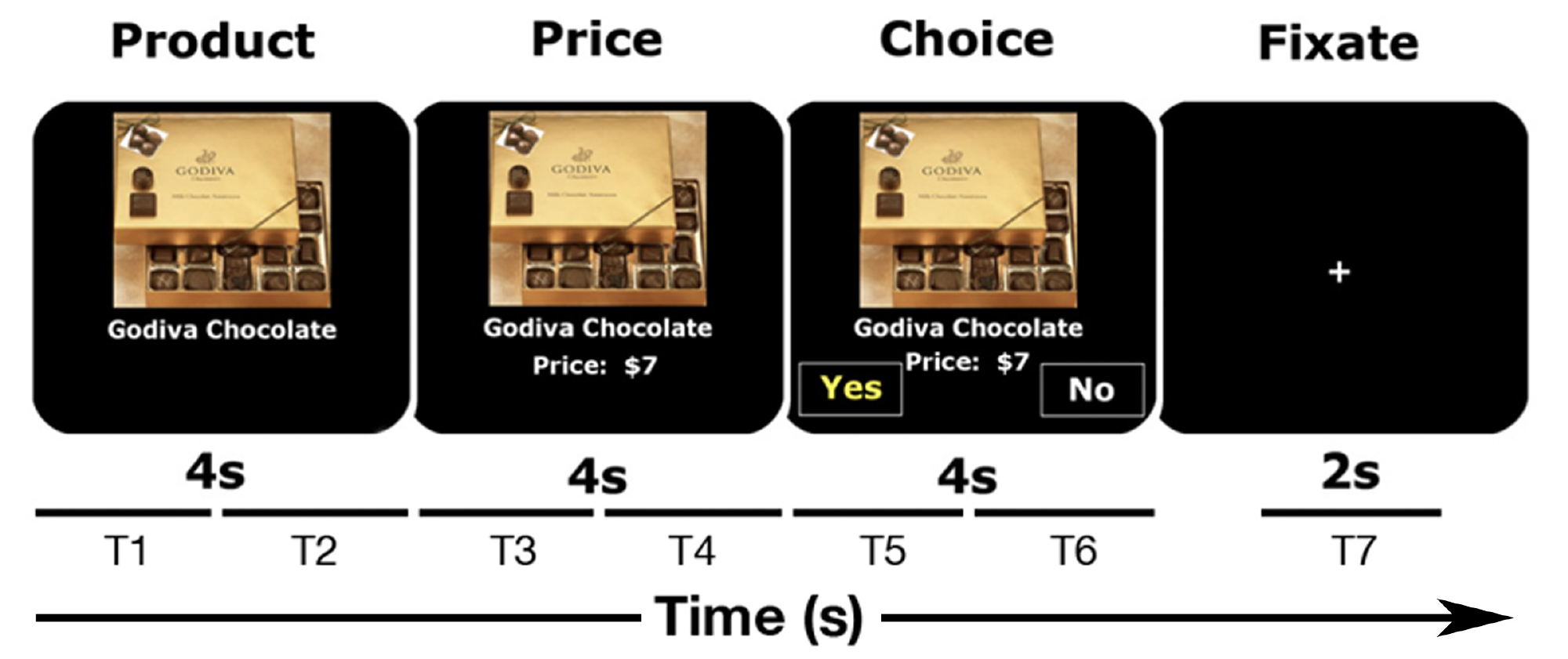}}
\par\end{centering}

\textcolor{black}{\medskip{}
}

\begin{spacing}{0.5}
\raggedright{}\textbf{\textcolor{black}{\footnotesize Figure 3.}}\textcolor{black}{\footnotesize{}
Save Holdings, or Purchase (SHOP) task trial structure. Images represent
what the subject saw, bars represent 2 second TRs (T1-T7). Subjects
saw a labeled product (product period; 4 s, 2 TRs), saw the product\textquoteright{}s
price (price period; 4 s, 2 TRs), and then chose either to purchase
the product or not (by selecting either \textquoteleft{}\textquoteleft{}yes\textquoteright{}\textquoteright{}
or \textquoteleft{}\textquoteleft{}no\textquoteright{}\textquoteright{}
presented randomly on the right or left side of the screen; choice
period; 4 s, 2 TRs), before fixating on a crosshair (2 s, 1 TR) prior
to the onset of the next trial. }\end{spacing}
\end{figure}
\textcolor{black}{{} Data from 25 healthy right-handed subjects were
analyzed (Knutson et al., 2007). Along with the typical magnetic resonance
exclusions (e.g., metal in the body), subjects were screened for psychotropic
drugs, cardiac drugs, ibuprofen, substance abuse in the past month,
and history of psychiatric disorders (DSM IV Axis I) prior to collecting
informed consent. Subjects were paid \$20.00 per hour for participating
and also received \$40.00 in cash to spend on products. Of 40 total
subjects, 6 subjects who purchased fewer than four items per session
(i.e., $<10\%$) were excluded due to insufficient data to fit, 8
subjects who moved excessive amounts (i.e., $>2$ mm between whole
brain acquisitions) were excluded, and one subject's original fMRI
data could not be recovered and so was omitted, yielding the final
total of 25 subjects included in the analysis.}

\textcolor{black}{While being scanned, subjects participated in a
\textquotedbl{}Save Holdings Or Purchase\textquotedbl{} (SHOP) task
(Figure 3). During each task trial, subjects saw a labeled product
(product period; 4 sec), saw the product's price (price period; 4
sec), and then chose either to purchase the product or not (by selecting
either \textquotedbl{}yes\textquotedbl{} or \textquotedbl{}no\textquotedbl{}
presented randomly on the right or left side of the screen; choice
period; 4 s), before fixating on a crosshair (2 s) prior to the onset
of the next trial (see Figure 3). }

\textcolor{black}{Each of 80 trials featured a different product.
Products were pre-selected to have above-median attractiveness, as
rated by a similar sample in a pilot study. While products ranged
in retail price from \$8.00-\$80.00, the associated prices that subjects
saw in the scanner were discounted down to 25\% of retail value to
encourage purchasing. Therefore the cost of each product during the
experiment ranged from \$2.00 to \$20.00. Consistent with pilot findings,
this led subjects to purchase 30\% of the products on average, generating
sufficient instances of purchasing to fit. }

\textcolor{black}{To ensure subjects' engagement in the task, two
trials were randomly selected after scanning to count \textquotedbl{}for
real\textquotedbl{}. If subjects had chosen to purchase the product
presented during the randomly selected trial, they paid the price
that they had seen in the scanner from their \$40.00 endowment and
were shipped the product within two weeks. If not, subjects kept their
\$40.00 endowment. Based on these randomly drawn trials, seven of
twenty-five subjects (28\%) were actually shipped products. }

\textcolor{black}{Subjects were instructed in the task and tested
for comprehension prior to entering the scanner. During scanning,
subjects chose from 40 items twice and then chose from a second set
of 40 items twice (80 items total), with each set presented in the
same pseudo-random order (item sets were counterbalanced across subjects).
We consider only data from the first time each item was presented
here (see \citet{Grosenick:2008p2789} for a comparison between first
and second presentations). After scanning, subjects rated each product
in terms of how much they would like to own it and what percentage
of the retail price they would be willing to pay for it. Then, two
trials were randomly drawn to count \textquotedbl{}for real\textquotedbl{},
and subjects received the outcome of each of the drawn trials.}
\end{singlespace}

A second validation sample included 17 healthy right-handed subjects
\citep{Karmarkar:2012}. These subjects passed the same screening,
inclusion, and exclusion criteria. Of an original sample of 24, 6
subjects purchased fewer than four items per session, and one showed
excessive motion. These subjects were excluded from analyses, as before.
Subjects also received the same payment and underwent the same scanning
and experimental procedures. Importantly, however, subjects were different
individuals who were exposed to different products, and were scanned
more than three years after the original study.\textcolor{black}{\medskip{}
}

\begin{singlespace}

\subsubsection{\textcolor{black}{Image acquisition}}
\end{singlespace}

\textcolor{black}{\medskip{}
}

\begin{singlespace}
\textcolor{black}{Functional images were acquired with a 1.5 T General
Electric MRI scanner using a standard birdcage quadrature head coil.
Twenty-four 4-mm-thick slices (in-plane resolution 3.75 X 3.75 mm,
no gap) extended axially from the midpons to the top of the skull,
providing whole-brain coverage and adequate spatial resolution of
subcortical regions of interest (e.g., midbrain, NAcc, OFC). Whole-brain
functional scans were acquired with a T2{*}-sensitive spiral in-/out-
pulse sequence (TR=2 s, TE=40 ms, flip=90), which minimizes signal
dropout at the base of the brain \citep{GloverLaw}. High-resolution
structural scans were also acquired to facilitate localization and
coregistration of functional data, using a T1-weighted spoiled grass
sequence (TR=100 ms, TE=7 ms, flip=90).}
\end{singlespace}

\textcolor{black}{\medskip{}
}

\begin{singlespace}

\subsubsection{\textcolor{black}{Preprocessing}}
\end{singlespace}

\textcolor{black}{\medskip{}
}

\begin{singlespace}
\textcolor{black}{After reconstruction, preprocessing was conducted
using Analysis of Functional Neural Images (AFNI) software \citep{CoxAFNI}.
For all functional images, voxel time-series were sinc interpolated
to correct for non-simultaneous slice acquisition within each volume,
concatenated across runs, corrected for motion, and normalized to
percent signal change with respect to the voxel mean for the entire
task. For further preprocessing details see \citep{Grosenick:2008p2789}.
Given that spatial blur would artificially increase correlations between
variables for the voxel-wise analysis, we used data with no spatial
blur and a temporal high pass filter for all analyses. Note that in
general, smoothing before running analyses will compound the problems
with correlation mentioned above, resulting in {}``rougher'' (high-frequency)
coefficients overall (see discussion in \citep{Hastie:1995p2589}). }

\textcolor{black}{Spatiotemporal data were arranged as in previous
spatiotemporal analyses \citep{MouraoMiranda:2007p2565}. Specifically,
data was arranged as an $n\times p$ data matrix $X$ with $n$ corresponding
to the number of trial observations on the $p$ input variables, each
of which was a particular voxel at a particular time point. This yielded
26,630 voxels taken at 7 time points (each taken every 2 seconds),
yielding a total of $p=186,410$ input input features per trial. Altogether,
the data used for training and test from \citep{Knutson2007} included
$n=1,882$ trials across the 25 subjects. The validation sample from
\citep{Karmarkar:2012} included $n=322$ trials across the 17 subjects.
In the first case (training and testing on the \citet{Knutson2007}
data), the number of 'buy' trials were upsampled to match the number
of 'not buy' trials in order to efficiently use the data when fitting
the models. In the out-of-sample (OOS) validation on the \citet{Karmarkar:2012}
data, however, the number of 'not buy' trials were downsampled to
match the smaller number of 'buy' trials in order to be more conservative
in estimating the out-of-sample accuracy (and related $p$-values).}
\end{singlespace}

\textcolor{black}{\medskip{}
}

\begin{singlespace}
\textcolor{black}{}
\begin{table}[t]
\begin{centering}
\textcolor{black}{\caption{Median classification accuracy and parameters for SPDA and SVM classifiers
fit with Leave-5-Subjects-Out (L5SO) cross validation.}
}
\par\end{centering}

\textcolor{black}{\medskip{}
}

\begin{centering}
\textcolor{black}{}%
\begin{tabular}{lcccc||r@{\extracolsep{0pt}.}lr@{\extracolsep{0pt}.}lr@{\extracolsep{0pt}.}lr@{\extracolsep{0pt}.}lr@{\extracolsep{0pt}.}l}
\multicolumn{5}{c}{Classification Accuracy} & \multicolumn{10}{c}{Model Type}\tabularnewline
\hline 
\hline 
\multirow{2}{*}{\textcolor{black}{\scriptsize Method}} & \multirow{2}{*}{\textcolor{black}{\scriptsize Training}} & \multirow{2}{*}{\textcolor{black}{\scriptsize Test}} & \multirow{2}{*}{{\scriptsize OOS }} & \multirow{2}{*}{{\scriptsize p-value$^{\dagger}$}} & \multicolumn{2}{c}{\textcolor{black}{\scriptsize Sparse}} & \multicolumn{2}{c}{{\scriptsize Tikhonov}} & \multicolumn{2}{c}{\textcolor{black}{\scriptsize Structured}} & \multicolumn{2}{c}{\textcolor{black}{\scriptsize Robust}} & \multicolumn{2}{c}{\textcolor{black}{\scriptsize Adaptive}}\tabularnewline
 &  &  &  &  & \multicolumn{2}{c}{\textcolor{black}{\scriptsize ($\lambda_{1}$)}} & \multicolumn{2}{c}{{\scriptsize ($\lambda_{2}$)}} & \multicolumn{2}{c}{\textcolor{black}{\scriptsize ($\lambda_{G}$)}} & \multicolumn{2}{c}{\textcolor{black}{\scriptsize ($\delta$)}} & \multicolumn{2}{c}{\textcolor{black}{\scriptsize ($\lambda_{1}^{*}$)}}\tabularnewline
\hline 
\hline 
 &  &  &  &  & \multicolumn{2}{c}{} & \multicolumn{2}{c}{} & \multicolumn{2}{c}{} & \multicolumn{2}{c}{} & \multicolumn{2}{c}{}\tabularnewline
\textcolor{black}{\scriptsize Linear SVM$^{1}$} & \textcolor{black}{\scriptsize $97.9\%$} & \textcolor{black}{\scriptsize $\mbox{71.0\%}$} & {\scriptsize $65.8\%$} & {\scriptsize $2.7\times10^{-8}$} & \multicolumn{2}{c}{} & \multicolumn{2}{c}{\textcolor{black}{\scriptsize $^{\dagger\dagger}3.8\times10^{-6}$}} & \multicolumn{2}{c}{} & \multicolumn{2}{c}{\textcolor{black}{\scriptsize $\checkmark$}} & \multicolumn{2}{c}{}\tabularnewline
 &  &  &  &  & \multicolumn{2}{c}{} & \multicolumn{2}{c}{} & \multicolumn{2}{c}{} & \multicolumn{2}{c}{} & \multicolumn{2}{c}{}\tabularnewline
\textcolor{black}{\scriptsize Lasso$^{2}$} & \textcolor{black}{\scriptsize $\mathbf{98.8}\%$} & \textcolor{black}{\scriptsize $68.5\%$} & {\scriptsize $58.4\%$} & {\scriptsize $0.003$} & \multicolumn{2}{c}{\textcolor{black}{\scriptsize $33$}} & \multicolumn{2}{c}{} & \multicolumn{2}{c}{} & \multicolumn{2}{c}{} & \multicolumn{2}{c}{}\tabularnewline
 &  &  &  &  & \multicolumn{2}{c}{} & \multicolumn{2}{c}{} & \multicolumn{2}{c}{} & \multicolumn{2}{c}{} & \multicolumn{2}{c}{}\tabularnewline
\textcolor{black}{\scriptsize Elastic Net$^{3}$} & \textcolor{black}{\scriptsize $90.4\%$} & \textcolor{black}{\scriptsize $72.5\%$ } & {\scriptsize $64.3\%$} & {\scriptsize $3.3\times10^{-7}$} & \multicolumn{2}{c}{\textcolor{black}{\scriptsize $54$}} & \multicolumn{2}{c}{\textcolor{black}{\scriptsize $10000$}} & \multicolumn{2}{c}{} & \multicolumn{2}{c}{} & \multicolumn{2}{c}{}\tabularnewline
 &  &  &  &  & \multicolumn{2}{c}{} & \multicolumn{2}{c}{} & \multicolumn{2}{c}{} & \multicolumn{2}{c}{} & \multicolumn{2}{c}{}\tabularnewline
\hline 
 &  &  &  &  & \multicolumn{2}{c}{} & \multicolumn{2}{c}{} & \multicolumn{2}{c}{} & \multicolumn{2}{c}{} & \multicolumn{2}{c}{}\tabularnewline
\textcolor{black}{\scriptsize GraphNet$^{4}$ (GN)} & \textcolor{black}{\scriptsize $86.9\%$} & \textcolor{black}{\scriptsize $73.7\%$ } & {\scriptsize $64.6\%$} & {\scriptsize $1.8\times10^{-7}$} & \multicolumn{2}{c}{\textcolor{black}{\scriptsize $68$}} & \multicolumn{2}{c}{\textcolor{black}{\scriptsize $1000$}} & \multicolumn{2}{c}{\textcolor{black}{\scriptsize $100$}} & \multicolumn{2}{c}{} & \multicolumn{2}{c}{}\tabularnewline
 &  &  &  &  & \multicolumn{2}{c}{} & \multicolumn{2}{c}{} & \multicolumn{2}{c}{} & \multicolumn{2}{c}{} & \multicolumn{2}{c}{}\tabularnewline
\textcolor{black}{\scriptsize Robust GN (RGN)} & \textcolor{black}{\scriptsize $86.8\%$} & \textbf{\textcolor{black}{\scriptsize 74.5$\%$}} & {\scriptsize $64.9\%$} & {\scriptsize $1.8\times10^{-7}$} & \multicolumn{2}{c}{\textcolor{black}{\scriptsize $43$}} & \multicolumn{2}{c}{\textcolor{black}{\scriptsize $100$}} & \multicolumn{2}{c}{\textcolor{black}{\scriptsize $100$}} & \multicolumn{2}{c}{\textcolor{black}{\scriptsize $0.3$}} & \multicolumn{2}{c}{}\tabularnewline
 &  &  &  &  & \multicolumn{2}{c}{} & \multicolumn{2}{c}{} & \multicolumn{2}{c}{} & \multicolumn{2}{c}{} & \multicolumn{2}{c}{}\tabularnewline
\textcolor{black}{\scriptsize RGN + temporal} & \textcolor{black}{\scriptsize $96.5\%$} & \textcolor{black}{\scriptsize $73.8\%$} & {\scriptsize $63.0\%$} & {\scriptsize $5.7\times10^{-6}$} & \multicolumn{2}{c}{\textcolor{black}{\scriptsize $42$}} & \multicolumn{2}{c}{\textcolor{black}{\scriptsize $1000$}} & \multicolumn{2}{c}{\textcolor{black}{\scriptsize $10$}} & \multicolumn{2}{c}{\textcolor{black}{\scriptsize $0.5$}} & \multicolumn{2}{c}{}\tabularnewline
 &  &  &  &  & \multicolumn{2}{c}{} & \multicolumn{2}{c}{} & \multicolumn{2}{c}{} & \multicolumn{2}{c}{} & \multicolumn{2}{c}{}\tabularnewline
\textcolor{black}{\scriptsize Adaptive RGN } & \textcolor{black}{\scriptsize $91.4\%$} & \textcolor{black}{\scriptsize $73.8\%$} & {\scriptsize $\mathbf{67.1}\%$} & {\scriptsize $8.6\times10^{-10}$} & \multicolumn{2}{c}{\textcolor{black}{\scriptsize $50$}} & \multicolumn{2}{c}{\textcolor{black}{\scriptsize $10000$}} & \multicolumn{2}{c}{\textcolor{black}{\scriptsize $100$}} & \multicolumn{2}{c}{\textcolor{black}{\scriptsize $0.4$}} & \multicolumn{2}{c}{\textcolor{black}{\scriptsize $0.01$}}\tabularnewline
 &  &  &  &  & \multicolumn{2}{c}{} & \multicolumn{2}{c}{} & \multicolumn{2}{c}{} & \multicolumn{2}{c}{} & \multicolumn{2}{c}{}\tabularnewline
\textcolor{black}{\scriptsize ARGN + temporal} & \textcolor{black}{\scriptsize $90.8\%$} & \textcolor{black}{\scriptsize $73.5\%$} & {\scriptsize $66.8\%$} & {\scriptsize $1.8\times10^{-9}$} & \multicolumn{2}{c}{\textcolor{black}{\scriptsize $40$}} & \multicolumn{2}{c}{\textcolor{black}{\scriptsize $1000$}} & \multicolumn{2}{c}{\textcolor{black}{\scriptsize $100$}} & \multicolumn{2}{c}{\textcolor{black}{\scriptsize $0.3$}} & \multicolumn{2}{c}{\textcolor{black}{\scriptsize $0.01$}}\tabularnewline
 &  &  &  &  & \multicolumn{2}{c}{} & \multicolumn{2}{c}{} & \multicolumn{2}{c}{} & \multicolumn{2}{c}{} & \multicolumn{2}{c}{}\tabularnewline
\textcolor{black}{\scriptsize Support Vector GN} & \textcolor{black}{\scriptsize $85.3\%$} & \textcolor{black}{\scriptsize $73.0\%$} & {\scriptsize $62.4\%$} & {\scriptsize $1.6\times10^{-5}$} & \multicolumn{2}{c}{\textcolor{black}{\scriptsize $120$}} & \multicolumn{2}{c}{\textcolor{black}{\scriptsize $1000$}} & \multicolumn{2}{c}{\textcolor{black}{\scriptsize $10$}} & \multicolumn{2}{c}{\textcolor{black}{\scriptsize $0.5$}} & \multicolumn{2}{c}{}\tabularnewline
 &  &  &  &  & \multicolumn{2}{c}{} & \multicolumn{2}{c}{} & \multicolumn{2}{c}{} & \multicolumn{2}{c}{} & \multicolumn{2}{c}{}\tabularnewline
\hline 
\hline 
 &  &  &  & \multicolumn{1}{c}{} & \multicolumn{2}{c}{} & \multicolumn{2}{c}{} & \multicolumn{2}{c}{} & \multicolumn{2}{c}{} & \multicolumn{2}{c}{}\tabularnewline
\end{tabular}
\par\end{centering}

\textcolor{black}{\scriptsize $^{1}$\citep{Cortes1995},$^{2}$\citep{Tibs1996},
$^{3}$\citep{ZouHastie}, $^{4}$\citep{HBM2009}. OOS is short for
{}``Out-Of-Sample''. Chance level is 50\%. $\dagger$ p-value is
calculated for the out-of-sample accuracy using an exact test for
the probability of success in a Bernoulli experiment with $n=322$
trials with success probability of 0.5. $\dagger\dagger$ This is
the $C$ parameter for the SVM. $\checkmark$ The linear SVM is robust
as a result of its hinge loss function, which does not have a parameter
$\delta$ associated with it.}
\end{table}

\end{singlespace}

\textcolor{black}{}
\begin{table}[t]
\begin{centering}
\textcolor{black}{\caption{Median classification accuracy and parameters for SPDA and SVM classifiers
fit with Leave-One-Subject-Out (LOSO) cross validation.}
}
\par\end{centering}

\textcolor{black}{\medskip{}
}

\begin{centering}
\textcolor{black}{}%
\begin{tabular}{lcccc||ccccc}
\multicolumn{5}{c}{Classification Accuracy} & \multicolumn{5}{c}{Model Type}\tabularnewline
\hline 
\hline 
\multirow{2}{*}{\textcolor{black}{\scriptsize Method}} & \multirow{2}{*}{\textcolor{black}{\scriptsize Training}} & \multirow{2}{*}{\textcolor{black}{\scriptsize Test}} & \multirow{2}{*}{{\scriptsize OOS }} & \multirow{2}{*}{{\scriptsize p-value$^{\dagger}$}} & \textcolor{black}{\scriptsize Sparse} & {\scriptsize Tikhonov} & \textcolor{black}{\scriptsize Structured} & \textcolor{black}{\scriptsize Robust} & \textcolor{black}{\scriptsize Adaptive}\tabularnewline
 &  &  &  &  & \textcolor{black}{\scriptsize ($\lambda_{1}$)} & {\scriptsize ($\lambda_{2}$)} & \textcolor{black}{\scriptsize ($\lambda_{G}$)} & \textcolor{black}{\scriptsize ($\delta$)} & \textcolor{black}{\scriptsize ($\lambda_{1}^{*}$)}\tabularnewline
\hline 
\hline 
 &  &  &  &  &  &  &  &  & \tabularnewline
\textcolor{black}{\scriptsize Linear SVM$^{1}$} & \textcolor{black}{\scriptsize $\mathbf{91.6}\%$} & \textcolor{black}{\scriptsize $\mbox{68.8\%}$} & {\scriptsize $65.2\%$} & {\scriptsize $9.7\times10^{-8}$} &  & \textcolor{black}{\scriptsize $^{\dagger\dagger}7.6\times10^{-6}$} &  & \textcolor{black}{\scriptsize $\checkmark$} & \tabularnewline
 &  &  &  &  &  &  &  &  & \tabularnewline
\textcolor{black}{\scriptsize Lasso$^{2}$} & \textcolor{black}{\scriptsize $90.5\%$} & \textcolor{black}{\scriptsize $68.8\%$} & {\scriptsize $61.2\%$} & {\scriptsize $7.1\times10^{-5}$} & \textcolor{black}{\scriptsize $63$} &  &  &  & \tabularnewline
 &  &  &  &  &  &  &  &  & \tabularnewline
\textcolor{black}{\scriptsize Elastic Net$^{3}$} & \textcolor{black}{\scriptsize $90.8\%$} & \textcolor{black}{\scriptsize $70.0\%$ } & {\scriptsize $63.0\%$} & {\scriptsize $5.7\times10^{-6}$} & \textcolor{black}{\scriptsize $61$} & \textcolor{black}{\scriptsize $1000$} &  &  & \tabularnewline
 &  &  &  &  &  &  &  &  & \tabularnewline
\hline 
 &  &  &  &  &  &  &  &  & \tabularnewline
\textcolor{black}{\scriptsize GraphNet$^{4}$ (GN)} & \textcolor{black}{\scriptsize $87.5\%$} & \textcolor{black}{\scriptsize $71.3\%$ } & {\scriptsize $67.7\%$} & {\scriptsize $4.1\times10^{-10}$} & \textcolor{black}{\scriptsize $54$} & \textcolor{black}{\scriptsize $10000$} & \textcolor{black}{\scriptsize $1000$} &  & \tabularnewline
 &  &  &  &  &  &  &  &  & \tabularnewline
\textcolor{black}{\scriptsize Robust GN (RGN)} & \textcolor{black}{\scriptsize $83.8\%$} & \textcolor{black}{\scriptsize $72.5\%$} & {\scriptsize $67.4\%$} & {\scriptsize $4.1\times10^{-10}$} & \textcolor{black}{\scriptsize $25$} & \textcolor{black}{\scriptsize $10$} & \textcolor{black}{\scriptsize $100$} & \textcolor{black}{\scriptsize $0.2$} & \tabularnewline
 &  &  &  &  &  &  &  &  & \tabularnewline
\textcolor{black}{\scriptsize RGN + temporal} & \textcolor{black}{\scriptsize $83.8\%$} & \textcolor{black}{\scriptsize $72.5\%$} & {\scriptsize $67.1\%$} & {\scriptsize $8.6\times10^{-10}$} & \textcolor{black}{\scriptsize $55$} & \textcolor{black}{\scriptsize $100$} & \textcolor{black}{\scriptsize $1000$} & \textcolor{black}{\scriptsize $0.6$} & \tabularnewline
 &  &  &  &  &  &  &  &  & \tabularnewline
\textcolor{black}{\scriptsize Adaptive RGN } & \textcolor{black}{\scriptsize $85.4\%$} & \textcolor{black}{\scriptsize $72.5\%$} & {\scriptsize $\mathbf{69.8}\%$} & {\scriptsize $1.7\times10^{-12}$} & \textcolor{black}{\scriptsize $20$} & \textcolor{black}{\scriptsize $10$} & \textcolor{black}{\scriptsize $1000$} & \textcolor{black}{\scriptsize $0.2$} & \textcolor{black}{\scriptsize $0.01$}\tabularnewline
 &  &  &  &  &  &  &  &  & \tabularnewline
\textcolor{black}{\scriptsize ARGN + temporal} & \textcolor{black}{\scriptsize $88.3\%$} & \textbf{\textcolor{black}{\scriptsize 73.8$\%$}} & {\scriptsize $68.9\%$} & {\scriptsize $2.0\times10^{-11}$} & \textcolor{black}{\scriptsize $30$} & \textcolor{black}{\scriptsize $1000$} & \textcolor{black}{\scriptsize $100$} & \textcolor{black}{\scriptsize $0.2$} & \textcolor{black}{\scriptsize $0.01$}\tabularnewline
 &  &  &  &  &  &  &  &  & \tabularnewline
\textcolor{black}{\scriptsize Support Vector GN} & \textcolor{black}{\scriptsize $89.5\%$} & \textbf{\textcolor{black}{\scriptsize 73.8$\%$}} & {\scriptsize $65.2\%$} & {\scriptsize $9.7\times10^{-8}$} & \textcolor{black}{\scriptsize $84$} & \textcolor{black}{\scriptsize $100$} & \textcolor{black}{\scriptsize $100$} & \textcolor{black}{\scriptsize $0.5$} & \tabularnewline
 &  &  &  &  &  &  &  &  & \tabularnewline
\hline 
\hline 
 &  &  &  & \multicolumn{1}{c}{} &  &  &  &  & \tabularnewline
\end{tabular}
\par\end{centering}

\textcolor{black}{\scriptsize $^{1}$\citep{Cortes1995},$^{2}$\citep{Tibs1996},
$^{3}$\citep{ZouHastie}, $^{4}$\citep{HBM2009}. OOS is short for
{}``Out-Of-Sample''. Chance level is 50\%. $\dagger$ p-value is
calculated for the out-of-sample accuracy using an exact test for
the probability of success in a Bernoulli experiment with $n=322$
trials with chance level at 50\%. $\dagger\dagger$ This is the $C$
parameter for the SVM. $\checkmark$ The linear SVM is robust as a
result of its hinge loss function, which does not have a parameter
$\delta$ associated with it.}
\end{table}

\textcolor{black}{}
\begin{figure}[H]
\begin{spacing}{0.5}
\begin{raggedright}
\textcolor{black}{\includegraphics[scale=0.8]{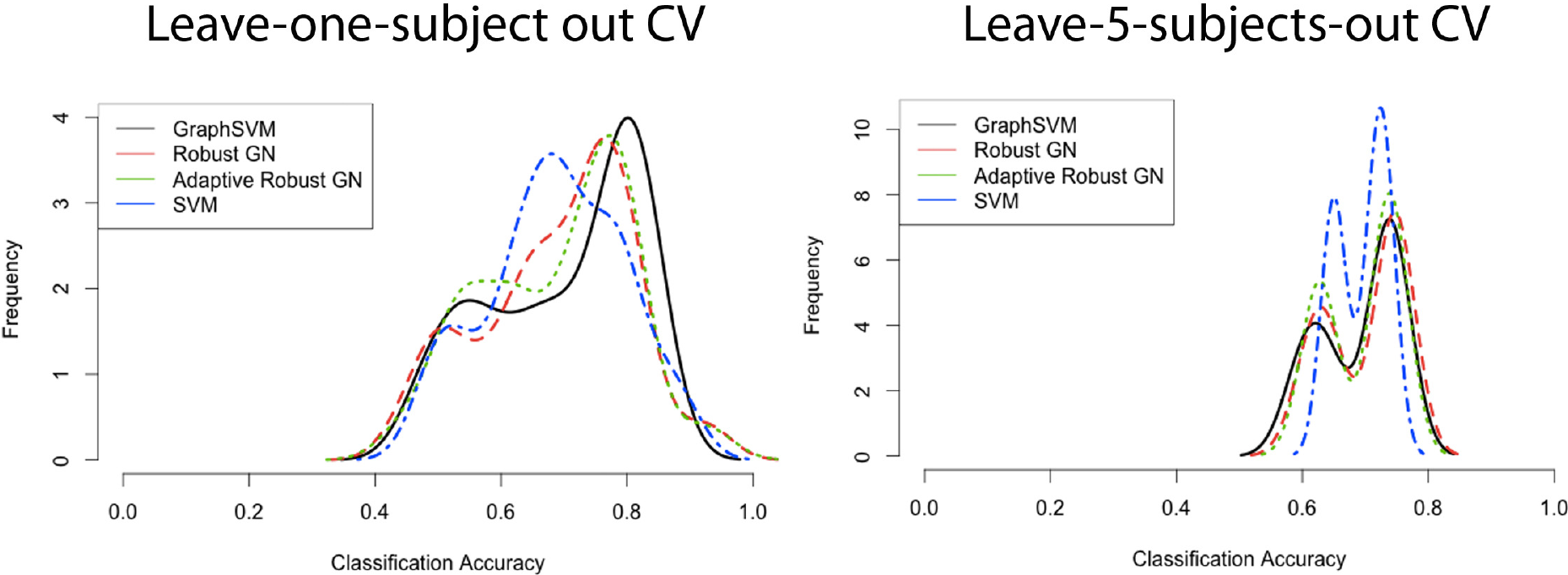}\medskip{}
}
\par\end{raggedright}

\raggedright{}\textbf{\textcolor{black}{\footnotesize Figure 4. }}\textcolor{black}{\footnotesize (Left)
Smoothed histogram densities of leave-one-subject out (LOSO) accuracy
rates on test data. Models were fit to all subjects except one, and
then tested on the held-out subject. This was done for all subjects
and smoothed histograms of these rates were calculated for the best
fitting models. (Right) The same procedure was repeated, but leaving
5 subjects out at a time for a total of 25 cross-validation folds.
Both plots show some bi-modality suggestive of different underlying
groups.}\end{spacing}
\end{figure}

\pagebreak{}

\section{Results}

\begin{singlespace}

\subsection{\textcolor{black}{Classification rates }}
\end{singlespace}

\textcolor{black}{\medskip{}
}

\textcolor{black}{If neural substrates implicated in choice show invariance
across individuals, a method that successfully identifies and uses
these substrates to predict choice should generalize well across subjects.
We compared the GraphNet classifier accuracies with accuracies obtained
using linear SVM (where accuracy in this case is the ability to correctly
predict a subject's choices to purchase a product or not). In particular
we looked at generalization of fits to held-out {}``test'' sets
(consisting of subjects held out of a particular stage of the cross
validation procedure, but still present in other cross-validation
stages), and to out-of-sample (OOS) data (new data never used at any
stage of the model fitting) consisting of different subjects from
another study \citep{Karmarkar:2012}. Results and model parameters
for the GraphNet classifiers and linear SVM across the 25 subjects
from \citet{Knutson2007} ({}``Training'', {}``Test'') and 17
subjects from \citet{Karmarkar:2012} ({}``OOS'') are listed in
Tables 1 and 2, as well as a summary of each method's properties.
Models were fit using either leave-one-subject-out (LOSO, Table 1)
or leave-5-subjects-out (L5SO, Table 2) cross-validation, and both
training and test results are displayed to allow comparison of overfitting
on the training data versus the held-out test data. As cross-validation
is known to yield an overly optimistic estimate of the true classification
error rate \citep{Hastie:2009p2681}, model fits to the initial data
set ($n=25$; \citet{Knutson2007}) were tested on out-of-sample (OOS)
data ($n=17$; \citet{Karmarkar:2012}) collected more than three
years later using different subjects shown different products. These
out-of-sample results provide the most rigorous demonstration of fit
generalization to new data, adjusting for any over fitting by the
cross-validation procedure, and are the strongest evidence for invariance
in the neural representation of choice across subjects. The $p$-values
reported correspond to these out-of-sample accuracies on $n=322$
trials across the 17 new subjects.}

\begin{singlespace}

\subsubsection{\textcolor{black}{Generalization to held-out groups (L5SO cross-validation)}}
\end{singlespace}

\textcolor{black}{\medskip{}
}

\begin{singlespace}
\textcolor{black}{Best median training, median test, and out-of-sample
(OOS) rates are described for GraphNet classifiers fit over the grid
of parameters given in \eqref{eq:16} The linear SVM parameters are
also given in \eqref{eq:16}. Despite a more than 1000-fold increase
in the number of input features relative to earlier volume of interest
(VOI) analyses \citep{Grosenick:2008p2789}, whole-brain classifiers
performed significantly better than previous VOI-based predictions
fit to the same data \citep{Knutson2007,Grosenick:2008p2789}. Further,
among these whole-brain classifiers, adaptive and robust methods performed
best on out-of-sample data. SVGN performed similarly to the linear
SVM (but unlike linear SVM, yields structured, sparse coefficients
that aid interpretability). Further, Lasso and linear SVM tended to
overfit the training data more than the SPDA-GraphNet classifiers,
as evidenced by their higher training but lower test rates. Overall,
the Adaptive Robust GraphNet classifier showed the best out-of-sample
classification rate, with accuracy on new data of 67.1\% (for comparison,
the linear SVM accuracy was 65.8\%). Examining the distribution of
test classification rates across the 25 folds (25 sets leaving 5 subjects
out), Figure 4b shows that the linear SVM appears to have less variance
across test fits to held-out subjects. The marked non-normality of
these distributions is interesting, and motivated us to report median
rather than mean accuracy over cross-validation folds.}
\end{singlespace}

\textcolor{black}{\medskip{}
}

\begin{singlespace}

\subsubsection{\textcolor{black}{Generalization to held-out individuals (LOSO cross-validation)}}
\end{singlespace}

\textcolor{black}{\medskip{}
}

\begin{singlespace}
\textcolor{black}{In addition to the leave-5-subject out (L5SO) cross-validation,
we also ran leave-one-subject-out (LOSO) cross validation (i.e., using
the data from 24 subjects to predict results for each remaining subject).
Repeating this procedure for all subjects yielded one held-out classification
rate per subject, indicating how well the group fit generalized to
that subject. Repeating this for all subjects yielded one held-out
test rate per subject. This rate indicated how well the model fit
based on all but one subjects' data generalized to the held-out subject---a
measure of invariance across subjects as well as a quantity that may
be of interest in studies of individual differences. Figure 4a shows
smoothed histograms of the LOSO classification rates for the Robust
GraphNet, Adaptive Robust GraphNet, SVGN, and linear SVM classifiers.
Overall, the GraphNet classifiers outperform the linear SVM on LOSO
cross-validation across subjects. When the LOSO fits were used to
classify choice out-of-sample, the Adaptive Robust GraphNet classifier
again yielded the best performance, now at almost 70\% classification
accuracy. LOSO cross-validation appears to result in better OOS generalization
than L5SO cross-validation for this data. More important than the
improvement in classification performance, however, is the greater
interpretability of these methods. }
\end{singlespace}

\textcolor{black}{\medskip{}
}

\begin{singlespace}

\subsection{\textcolor{black}{Visualization and interpretation of coefficients
and parameters\medskip{}
}}
\end{singlespace}

\begin{singlespace}

\subsubsection{\textcolor{black}{Interpreting GraphNet coefficients}}
\end{singlespace}

\textcolor{black}{\medskip{}
}

\begin{singlespace}
\textcolor{black}{While GraphNet classifiers and linear SVM both classified
purchase choices successfully, the GraphNet-based classifiers produced
more interpretable results. Consistent with previous VOI-based analyses,
the GraphNet, Robust GraphNet classifier (Figure 5), and Adaptive
Robust GraphNet (Figure 5) classifiers all identified similar regions
to those chosen as VOIs \citep{Knutson2007}, with coefficients present
at the time points corresponding to peak discrimination in the VOI
time-series \citep{Knutson2007} and VOI classification \citep{Grosenick:2008p2789}.
In particular, nucleus accumbens (NAcc) activation began to positively
predict purchase choices at the time of product presentation, and
this prediction persisted throughout subsequent price presentation.
Medial prefrontal cortex (MPFC) and midbrain activation, on the other
hand, began to positively predict purchase choices at the onset of
price presentation (but not during previous product presentation).
Additionally---and not included in any previous findings---posterior
cingulate activation also began to robustly and positively predict
purchase choices during price presentation. Reassuringly, no regions'
activation predicted purchase choices during fixation presentation.
Interestingly, the best fits chose far more voxels that positively
predicted than negatively predicted purchasing.}
\end{singlespace}

\textcolor{black}{Together, these findings demonstrate that sparse,
structured, whole-brain methods like GraphNet can facilitate the discovery
of new behaviorally-relevant spatiotemporal neural activity patterns
that existing VOI-based methods miss, particularly when made robust
and adaptive. For example, given the temporal as well as spatial resolution
of the present design and data, it was possible to extend interpretation
of the model fit not only to where brain activity predicted purchasing
choices, but also to when and in what order, and to new regions not
chosen as VOIs in previous work emerged and improved the overall classification.
Thus, an investigator who knows when different events occurred (and
accounts for the lag and variation of the peak hemodynamic response)
can infer that different design components promoted eventual purchasing
choices by altering activity in specific regions. The ability of coefficient
vectors estimated from the \citet{Knutson2007} to accurately predict
choices of new subjects run on the SHOP task years later and shown
different products speaks both to the stability of the neural activity
related to the task across subjects and products, and to the quality
of the model. \medskip{}
}

\textcolor{black}{}
\begin{figure}
\begin{centering}
\textcolor{black}{\includegraphics[scale=0.75]{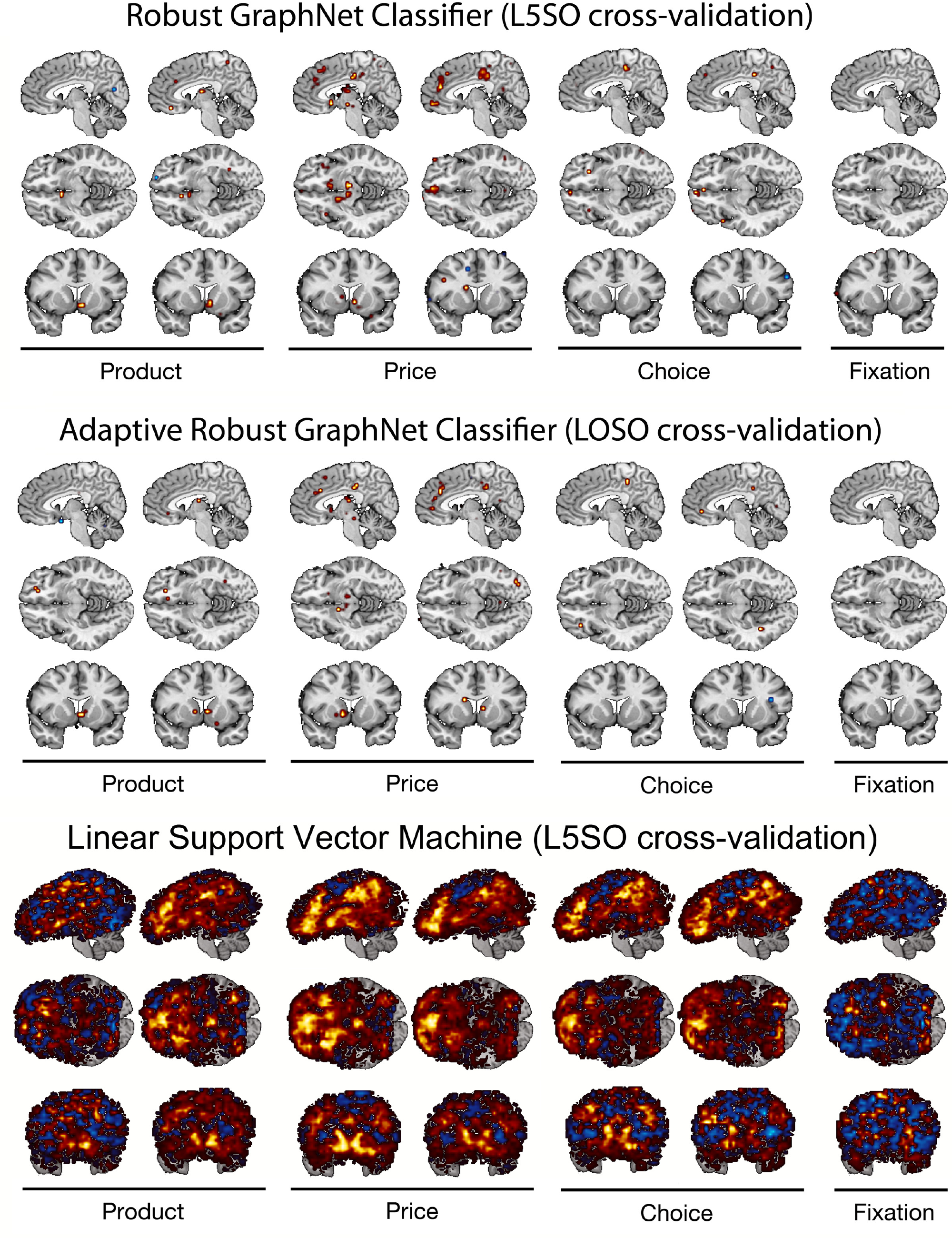}\medskip{}
}
\par\end{centering}

\begin{spacing}{0.5}
\raggedright{}\textbf{\textcolor{black}{\footnotesize Figure 5. }}\textcolor{black}{\footnotesize Whole-brain
classification results from the SHOP task (see Figure 3 for task structure).
(Top) Median coefficient maps from the best robust GraphNet classifier
(median test accuracy of 74.5\% over cross-validation folds and out-of-sample
accuracy of 64.9\%) fit using Leave-5-subject-out (L5SO) cross-validation
are shown at two time points for product, price, and choice periods,
as well as the fixation period. Warm colored coefficients denote areas
that predict purchasing a product, while cool-colored areas those
that predict not purchasing. The areas chosen by the robust GraphNet
classifier highlight regions suggested by previous studies including
the bilateral nucleus accumbens (NAcc) and the mesial prefrontal cortex
(MPFC) \citep{Knutson2007,Grosenick:2008p2789}, but also implicate
new regions including the anterior cingulate and and posterior cingulate
cortices. (Middle) Similar plots for the best adaptive robust GraphNet
classifier (median test accuracy of 72.5\% over cross-validation folds;
out-of-sample accuracy of 69.8\%) fit using leave-one-subject (LOSO)
cross-validation. Although the solution is sparser, the regions chosen
remain the same. (Bottom) Coefficients for the best linear SVM (median
test accuracy of 71\% over cross-validation folds; out-of-sample accuracy
of 65.8\%) fit using Leave-5-subject-out (L5SO) cross-validation for
comparison. }\end{spacing}
\end{figure}
\textcolor{black}{}
\begin{figure}
\begin{centering}
\textcolor{black}{\includegraphics[scale=0.8]{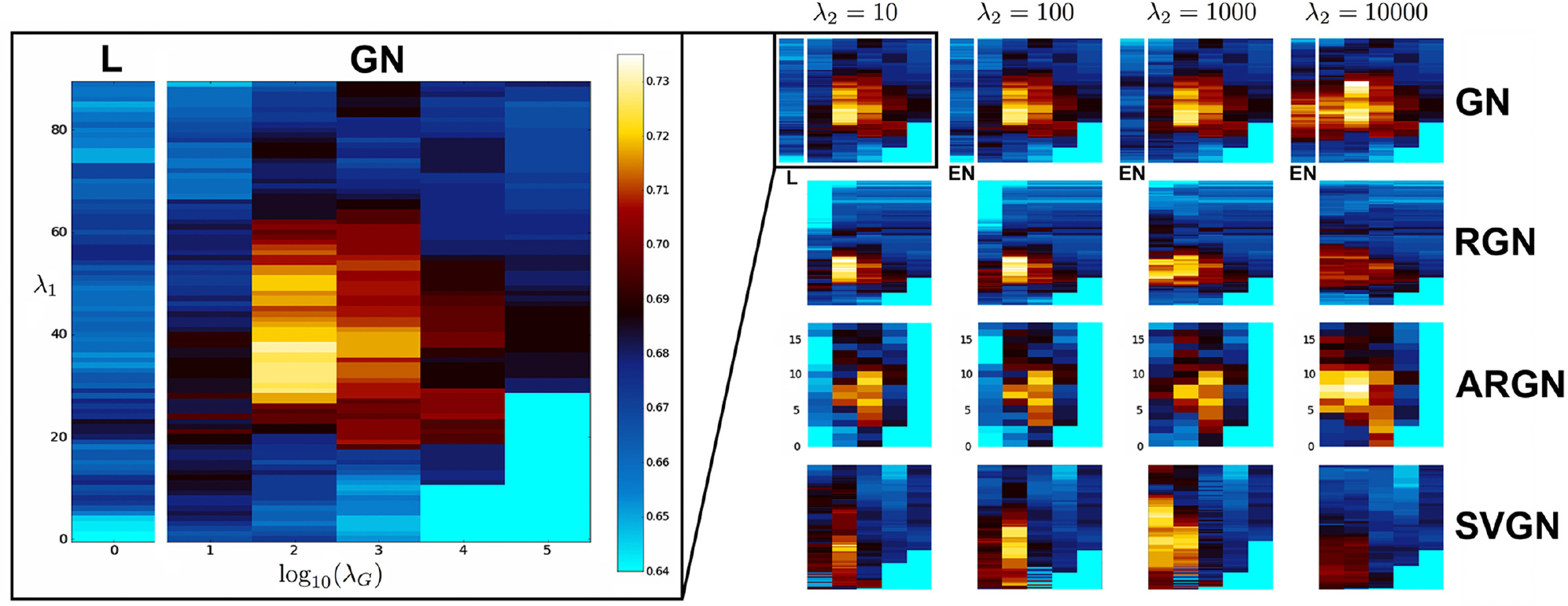}}
\par\end{centering}

\textcolor{black}{\medskip{}
}

\begin{spacing}{0.5}
\raggedright{}\textbf{\textcolor{black}{\footnotesize Figure 6. }}\textcolor{black}{\footnotesize Examples
of classification accuracy (test) plotted as a function of penalty
parameters. The blown up image on the left shows an image of the median
test accuracy rates for the GraphNet SPDA classifier (GN) as functions
of hyperparameters $\lambda_{1}$ and $\lambda_{G}$ (wit}\textbf{\textcolor{black}{\footnotesize h
$\lambda_{2}=0$}}\textcolor{black}{\footnotesize ). Warm colors indicate
median classification rates above 70\% (for L5SO cross-validation)
and cool colors median accuracy below 70\% (see color bar for scale).
The separate column (L) indicates the standard Lasso solution at }\textbf{\textcolor{black}{\footnotesize $\lambda_{G}=0$}}\textcolor{black}{\footnotesize .
There is a clear maxima at }\textbf{\textcolor{black}{\footnotesize $\lambda_{1}=40,\ \lambda_{G}=100$}}\textcolor{black}{\footnotesize .
The smaller images on the right show similar plots for the GraphNet
(GN), Robust GraphNet (RGN), Adaptive Robust GraphNet (ARGN), and
Support Vector GraphNet (SVGN) classifiers at four values of the graph
$G$ diagonal scale $\lambda_{2}$. Note the different scale on the
ARGN models. It is of some interest that all the plots are rather
slowly varying in the parameters and demonstrate significantly unimodal
peaks (neither of these need be the case).}\end{spacing}
\end{figure}

\begin{singlespace}

\subsubsection{\textcolor{black}{Interpreting GraphNet parameters}}
\end{singlespace}

\textcolor{black}{\medskip{}
}

\begin{singlespace}
\textcolor{black}{Figure 6 shows plots of median L5SO cross-validation
rates over the values $\{\lambda_{1},\lambda_{G},G\}$ \eqref{eq:15-1}
on which we fit the four GraphNet classifiers (other parameters are
set to the values shown in Table 1; plots for LOSO rates are similar).
In all cases, there is a region in the interior of the explored parameter
space $\{\lambda_{1},\lambda_{G},G\}$ on which the models empirically
perform best. In all cases this region involves both smoothing and
some level of sparsity, and the classifiers built with Lasso (L) and
Elastic Net (EN)---shown as separate bars for the GraphNet (GN) fits---underperform
relative to the sparse and smooth GraphNet classifiers on this data.
Comparison of the rates in Figure 6 suggests that a certain amount
of coefficient smoothness and inclusion of correlated variables in
the final fit is important for this data set, and that using a robust
loss function tightens the region of optimal parameter performance.}
\end{singlespace}

\textcolor{black}{\medskip{}
}

\begin{singlespace}

\section{\textcolor{black}{Discussion and Conclusions\medskip{}
}}
\end{singlespace}

\begin{singlespace}

\subsection{\textcolor{black}{Interpretable models for whole-brain spatiotemporal
fMRI}}
\end{singlespace}

\textcolor{black}{\medskip{}
}

\begin{singlespace}
\textcolor{black}{We sought to design and develop a novel classification
method for fMRI data that could fulfill several aims. First, the method
should deliver interpretable results for whole-brain data over multiple
time-points in the native data space. Second, the method should yield
classification accuracy (or goodness-of-fit) competitive with current
state-of-the-art multivariate methods. Third, the method should choose
relevant features in a principled and asymptotically consistent way
(i.e., it should include relevant features while excluding nuisance
parameters). Fourth, the method should accommodate flexible constraints
on model coefficients related to prior information (e.g., local smoothness,
connectivity). Fifth, the method should remain robust to outliers
in the data. Sixth, the method should generate coefficients with relatively
unbiased magnitudes (despite employing shrinkage methods to yield
sparsity). And seventh (and finally), the method should have the capacity
to detect a range of possible signals, from smooth and localized to
sparse and distributed. }

\textcolor{black}{The GraphNet-based methods presented here make a
first step toward meeting these desirable (and often competing) aims.
In particular, the Adaptive Robust GraphNet allows automatic variable
selection (\citet{Zou:2009p2991}), incorporation of prior information
in the form of a graph penalty, and yields minimally biased and asymptotically
consistent coefficient estimates as a result of adaptive reweighting.
Robust GraphNet methods can be applied to either regression or classification
settings (using Optimal Scoring), and generate classification rates
that compete favorably with state-of-the-art multivariate classifiers.
The tuning parameters $(\lambda_{1},\lambda_{G})$ and the graph $G$
allow for a diversity of sparse and smooth data, and the relationship
of model fits to these parameters provides information about the structure
of the detected signal.}

\textcolor{black}{Choice in the context of purchasing admittedly represents
only one application, and future validation on additional data sets
is necessary. However, in this context the GraphNet classifiers generalize
well to independent experiments involving purchasing (i.e. when fit
to new data collected years after the experiments originally used
to train the models, with different subjects and different products).
Adaptive Robust GraphNet methods showed the best out-of-sample generalization,
and generated parsimonious, interpretable models. It is worth noting
that the models that did best on the in-sample cross-validation test
folds were not best on out-of-sample data. This suggests that the
overfitting, or {}``optimism'', known to exist in cross-validation
\citep{Hastie:2009p2681} can effect models differently, and that
a true out-of-sample prediction is necessary to accurately assess
which models generalize best.}
\end{singlespace}

In summary, we have developed a family of robust, adaptive, and interpretable
methods that can be fit efficiently to large data sets over large
parameter grids. This method will allow investigators to search in
a data-driven fashion across the whole brain and multiple time points,
obviating the need for volume-of-interest based approaches in fMRI
classification and regression, and providing an effective alternative
to mass-univariate approaches for whole-brain analysis.

\textcolor{black}{\medskip{}
}

\begin{singlespace}

\subsection{\textcolor{black}{Application to SHOP task data}}
\end{singlespace}

\textcolor{black}{\medskip{}
}

\begin{singlespace}
\textcolor{black}{In the context of predicting human behavior from
brain data, the current whole brain methods offer clear advantages
over previous volume of interest based methods. In terms of classification
accuracy, previous work on the \citet{Knutson2007} data has resulted
in cross-validated test rates of 60\% (with a leave-one-out cross
validation using logistic regression on VOI-averaged data; see \citet{Knutson2007}
for details), and 67\% (with a $5\times2$ cross validation using
SPDA-Elastic Net on VOI voxel data; see \citet{Grosenick:2008p2789}
for details). Here, using the same preprocessing and data as in these
previous VOI-based approaches, but using GraphNet classifiers on whole-brain
data, we achieve test rates from $73.0-74.5\%$ for L5SO cross validation
and $71.3-73.8\%$ for LOSO cross validation. Further, out-of-sample
(OOS) rates for the GraphNet classifiers were $67.1\%$ (L5SO) and
and $69.8\%$ (LOSO). Thus, in this case, even out-of-sample rates
with GraphNet classifiers outperform in-sample cross validation test
rates on VOI-based classifiers---a considerable improvement. In taking
classification accuracy as a measure of goodness-of-fit, this indicates
that GraphNet classifiers result in better fits and improved generalization
relative to VOI methods, and suggests that the resulting coefficients
are a good representation of invariant features that discriminate
between choosing to purchase or not across subjects and products.}
\end{singlespace}

\textcolor{black}{Turning to examine the coefficients, we see that
the GraphNet classifiers reassuringly deliver findings consistent
with prior volume of interest based results \citep{Knutson2007,Grosenick:2008p2789},
replicating the observation that nucleus accumbens (NAcc) activation
begins to predict purchase choices during product presentation while
medial prefrontal cortical (MPFC) activation begins to predict purchase
choices during price presentation. It is also interesting to note
areas that were not included by previously applied methods, and might
not have been noticed if not for the whole-brain analysis (and which
might help account for the improved classification rates over previous
VOI analyses). }

\textcolor{black}{While one account posits that in the context of
fMRI, NAcc activation indexes gain predictions \citep{Knutson2001,Knutson2008},
an alternative account posits that NAcc activation instead indexes
gain prediction errors (e.g., \citealt{Hare2008}). To the extent
that gain predictions forecast future events while gain prediction
errors are adjustments of those forecasts after an error is detected,
the gain prediction account posits that NAcc activation in response
to products should predict subsequent purchase choices. Applied to
SHOP task data, the robust and adaptive robust GraphNet classifier
results clearly support the gain prediction functional account of
NAcc activity, since NAcc activation in response to products predicts
future choices to purchase, whereas MPFC activity does not. Instead,
MPFC activity predicts choice in response to later presented price
information, consistent with a value integration account (\citep{Knutson2005};
Figure 5). The GraphNet classifiers also revealed a previously unnoticed
result in which anterior and posterior cingulate activity clearly
predicts purchase choices at price presentation (Figure 5). Accounts
of cingulate function in the context of purchasing remain less developed
than similar accounts of NAcc and MPFC function. Nonetheless, this
result might be consistent with attentional and salience-based accounts
of posterior cingulate function \citep{McCoy2003}, and highlights
a region that deserves further investigation in the context of choice
prediction.}

\textcolor{black}{\medskip{}
}

\begin{singlespace}

\subsection{\textcolor{black}{Future directions}}
\end{singlespace}

\textcolor{black}{\medskip{}
}

\begin{singlespace}
\textcolor{black}{GraphNet methods can be further optimized, opening
new avenues for exploration. For instance, investigators might compare
graph constraints other than those related to just spatial-temporal
adjacency, including (1) weighted graphs derived from the data to
adapt to local smoothness, (2) cut-graphs derived from segmented brain
atlases that allow adjacent but functionally distinct regions to be
independently penalized, and (3) weighted graphs derived from structural
data, which would allow constraints on voxels adjacent on a connectivity
graph, rather than in space or time (see \citet{NgVaroquaux2012}
for a promising step in this direction). Further, investigators might
use the goodness-of-fit measure provided by  GraphNet to infer which
of a set of structural graphs best relates to functional data, or
to adaptively alter graph weights to explore structure in functional
data (in a Variational Bayes framework, for example). }
\end{singlespace}

\textcolor{black}{All of the methods considered above assume linear
relationships between input features and target variables. While this
assumption suffices in many cases, signal saturation effects alone
suggest that it might not faithfully mirror underlying physiological
signals. Nonlinear methods based on scatterplot smoothers have recently
been developed and shown to work well in combination with coordinate-wise
methods \citep{Ravikumar2009}, and previous work applying sparse
regression to features derived using factor analysis have yielded
promising results \citep{ESS,Wager2011}. Investigators might thus
combine nonlinear methods with sparse structured feature selection
methods \citep{Allen2011} to generate more flexible and accurate,
yet still interpretable, models of brain dynamics. Finally, we note
that because we are operating directly on voxels data, we are working
in the {}``native'' reconstructed 3D data space rather than on factors
derived from this data or on a dictionary of basis functions that
approximate features of the data (e.g., wavelets). Certainly, the
optimization scheme described here would also extend to solving problems
using features derived from the data, and it is an interesting direction
for future research to explore GraphNet penalties in these other contexts
and to compare GraphNet methods to existing regression and classification
methods that operate on lower dimensional embeddings or dictionary
representations of the data. Whether operating directly on the data
with sparse structured methods or on derived features is more appropriate
will depend on the application. The methods presented here demonstrate
that the former approach can be quite effective, and provides results
that are easily interpreted in the native data space. }

\pagebreak{}

\section*{\textcolor{black}{Appendix }}

\subsection*{\textcolor{black}{Robust GraphNet: coordinate-wise coefficient updates
using infimal convolution}}

\textcolor{black}{\medskip{}
}

\textcolor{black}{}
\begin{algorithm}[H]
\textcolor{black}{\caption{Robust GraphNet update using infimal convolution}
}
\begin{enumerate}
\item \textcolor{black}{Given a set of data and parameters $\Omega=\{X,y,\lambda_{1},\lambda_{G}\}$,
previous coefficient estimates $\widehat{\alpha}^{(r)},\widehat{\beta}^{(r)}$,
and $p\times p$ positive semidefinite constraint graph $G\in S_{+}^{p\times p}$,
let 
\begin{eqnarray*}
\widehat{\gamma}^{(r)} & = & [\widehat{\beta}^{(r)}\ \ \widehat{\alpha}^{(r)}]^{T}\\
Z & = & \left[X\ \ I_{n\times n}\right].
\end{eqnarray*}
}
\item \textcolor{black}{Choose coordinate $j$ using essentially cyclic
rule \citep{Tseng2001} and fix $\tilde{\gamma}=\{\gamma_{k}^{(r)}|k\neq j\}$,
$\tilde{Z}=Z._{\neq j},$$\tilde{\beta}=\{\beta_{k}^{(r)}|k\neq j\}$,
$\tilde{X}=X._{\neq j}$ .}
\item \textcolor{black}{Update $\widehat{\gamma}_{j}^{(r)}$ using
\[
\widehat{\gamma}_{j}^{(r+1)}\leftarrow\begin{cases}
\frac{S\left(Z._{j}^{T}(y-\tilde{Z}\tilde{\gamma})-(\lambda_{2}/2)\tilde{\gamma}^{T}(G'_{\neq j}.)._{j},\ \lambda_{1}/2\right)}{Z._{j}^{T}Z._{j}+\lambda_{G}G'_{jj}} & \text{if }j\in\{1,...,p\}\\
S\left((y-\tilde{Z}\tilde{\gamma})_{j},\ \lambda_{1}/2\right) & \text{if }j\in\{p+1,...,p+n\},
\end{cases}
\]
where $S(x,\lambda)$ is the element-wise soft-thresholding operator
in equation \eqref{eq:soft-thresh}. For adaptive version replace
$\lambda_{1}$ with $\lambda_{1}^{*}\widehat{w}_{j}$ in above update
(see section 2.1.5).}
\item \textcolor{black}{Repeat steps (1)-(3) cyclically for all $j\in\{1,\ldots,p+n\}$
until convergence (see discussion of convergence in \citet{Friedman:2007p36}). }
\item \textcolor{black}{Optional: rescale resulting estimates using method
from section 2.2.5.}\end{enumerate}
\end{algorithm}

\subsubsection*{\textcolor{black}{\medskip{}
}Derivation of updates in Algorithm 1\textcolor{black}{\medskip{}
}}

\textcolor{black}{For a particular coordinate $j$ , we are interested
in the estimates
\begin{eqnarray*}
\widehat{\gamma}_{j} & = & \underset{\gamma_{j}}{\text{argmin}}\ (1/2)\|y-\tilde{Z}\tilde{\gamma}-Z._{j}\gamma_{j}\|_{2}^{2}+\lambda_{G}\left(\tilde{\gamma}^{T}(G'_{\neq j}.)._{j}\gamma_{j}+G'_{jj}\gamma_{j}^{2}\right)+\lambda_{1}|\gamma_{j}|\ \text{if }j\in\{1,...,p\},\\
\widehat{\gamma}_{j} & = & \underset{\gamma_{j}}{\text{argmin}}\ (1/2)\|y-\tilde{Z}\tilde{\gamma}-Z._{j}\gamma_{j}\|_{2}^{2}+\delta|\gamma_{j}|\ \text{if }j\in\{p+1,...,p+n\}.
\end{eqnarray*}
By the arguments in section 2.4.1, this yields the coordinate-wise
updates
\begin{eqnarray*}
\widehat{\gamma}_{j} & \leftarrow & \frac{S\left(Z._{j}^{T}(y-\tilde{Z}\tilde{\gamma})-(\lambda_{2}/2)\tilde{\gamma}^{T}(G'_{-j}.)._{j},\ \lambda_{1}/2\right)}{Z._{j}^{T}Z._{j}+\lambda_{G}G'_{jj}}\ \ \text{if }j\in\{1,...,p\},\\
\widehat{\gamma}_{j} & \leftarrow & \frac{S\left(Z._{j}^{T}(y-\tilde{Z}\tilde{\gamma}),\ \lambda_{1}/2\right)}{Z._{j}^{T}Z._{j}}\ \ \text{if }j\in\{p+1,...,p+n\},
\end{eqnarray*}
where $S(x,\lambda)$ is the element-wise soft-thresholding operator
in equation \eqref{eq:soft-thresh}.}

\textcolor{black}{\medskip{}
}

\subsection*{\textcolor{black}{SVM-GraphNet classification: coordinate-wise coefficient
updates using infimal convolution}}

\textcolor{black}{\medskip{}
}

\textcolor{black}{}
\begin{algorithm}[H]
\textcolor{black}{\caption{SVM GraphNet classification update using infimal convolution}
}
\begin{enumerate}
\item \textcolor{black}{Given a set of data and parameters $\Omega=\{X,y,\lambda_{1},\lambda_{G}\}$,
previous coefficient estimates $\widehat{\alpha}^{(r)},\widehat{\beta}_{0}^{(r)},\widehat{\beta}^{(r)}$,
and $p\times p$ positive semidefinite constraint graph $G\in S_{+}^{p\times p}$,
let 
\begin{eqnarray*}
\widehat{\gamma}^{(r)} & = & [\widehat{\beta}_{0}^{(r)}\ \ \widehat{\beta}^{(r)}\ \ \widehat{\alpha}^{(r)}]^{T}\\
Z & = & \left[y^{T}[1_{n\times1}\ \ X]\ \ I_{n\times n}\right].
\end{eqnarray*}
}
\item \textcolor{black}{Choose coordinate $j$ using essentially cyclic
rule \citep{Tseng2001} and fix $\tilde{\gamma}=\{\gamma_{k}^{(r)}|k\neq j\}$,
$\tilde{Z}=Z._{\neq j},$$\tilde{\beta}=\{\beta_{k}^{(r)}|k\neq j\}$,
$\tilde{X}=X._{\neq j}$ .}
\item \textcolor{black}{Update $\widehat{\gamma}_{j}^{(r)}$ using
\[
\widehat{\gamma}_{j}^{(r+1)}\leftarrow\begin{cases}
\tilde{\gamma}^{T}\tilde{Z}1_{n\times1}+N(\gamma_{j}-1) & \text{if }j=0\\
\\
\frac{S\left((\tilde{Z}^{T}\tilde{\gamma}-1_{n\times1})^{T}X._{j}-(\lambda_{2}/2)\tilde{\beta}^{T}(G{}_{\neq j}.)._{j},\ \lambda_{1}/2\right)}{X._{j}^{T}X._{j}+\lambda_{G}G{}_{jj}} & \text{if }j\in\{1,...,p\}\\
\\
H\left((\tilde{Z}\tilde{\gamma})_{j}-1,\ \delta\right) & \text{if }j\in\{p+1,...,p+n\},
\end{cases}
\]
where $S(x,\lambda)$ is the element-wise soft-thresholding operator
and $H(x,\delta)$ is given in equation \eqref{eq: H}.}
\item \textcolor{black}{Repeat (1) -(3) cyclically for all $j\in\{1,\ldots,p+n\}$
until convergence (see discussion of convergence in \citet{Friedman:2007p36}).}
\item \textcolor{black}{Optional: rescale resulting estimates using method
from section 2.2.5.}\end{enumerate}
\end{algorithm}

\subsubsection*{\textcolor{black}{\medskip{}
}Derivation of updates in Algorithm 2\textcolor{black}{\medskip{}
}}

\textcolor{black}{Following the description of the SVM given in section
2.1.7, we can take the same approach used to derive the Robust GraphNet
estimates with the Support Vector GraphNet estimates of section 2.2.3,
which we can write as 
\begin{eqnarray*}
\widehat{\gamma} & = & \underset{\gamma}{\text{argmin}}\ (1/2\delta)\|1_{n\times1}-Z\gamma\|_{2}^{2}+\lambda_{G}\gamma_{\neq0}^{T}G'\gamma_{\neq0}+\sum_{j=0}^{p}w_{j}|\gamma_{j}|+\sum_{j=p+1}^{p+n}w_{j}\max(0,\gamma_{j})\\
\text{where} &  & Z=\left[y^{T}[1_{n\times1}\ \ X]\ \ I_{n\times n}\right],\ \ \gamma=[\beta_{0}\ \ \beta\ \ \alpha],\ \ w_{j}=\begin{cases}
0 & \text{if }j=0\\
\lambda_{1} & \text{if }j=1,..,p\\
1 & \text{if }j=p+1,...,p+n
\end{cases}\\
 &  & G'=\left[\begin{array}{lll}
0 & 0_{1\times p} & 0_{1\times n}\\
0_{p\times1} & G & 0_{1\times n}\\
0_{n\times1} & 0_{n\times1} & 0_{n\times n}
\end{array}\right]\in S_{+}^{(p+n+1)\times(p+n+1)}.
\end{eqnarray*}
During coordinate wise descent, only one of the separable penalty
functions has an {}``active'' variable per descent step. Letting
$h(\gamma_{j})=\max(0,\gamma_{j})$, we thus have 
\[
\widehat{\gamma}_{j}=\begin{cases}
\underset{\gamma_{j}}{\text{argmin}}\ (1/2\delta)\|1_{N\times1}-\tilde{Z}\tilde{\gamma}-Z._{j}\gamma_{j}\|_{2}^{2} & \text{if }j=0\\
\underset{\gamma_{j}}{\text{argmin}}\ (1/2\delta)\|1_{N\times1}-\tilde{Z}\tilde{\gamma}-Z._{j}\gamma_{j}\|_{2}^{2}+\lambda_{G}\left(\tilde{\gamma}^{T}(G'_{\neq j}.)._{j}\gamma_{j}+G'_{jj}\gamma_{j}^{2}\right)+\lambda_{1}|\gamma_{j}| & \text{if }j\in\{1,...,p\}\\
\underset{\gamma_{j}}{\text{argmin}}\ (1/2\delta)\|1_{N\times1}-\tilde{Z}\tilde{\gamma}-Z._{j}\gamma_{j}\|_{2}^{2}+h(\gamma_{j}) & \text{if }j\in\{p+1,...,p+n\}.
\end{cases}
\]
Then since 
\[
Z._{j}=\begin{cases}
1_{N\times1} & \text{if }j=0\\
X._{j} & \text{if }j\in\{1,...,p\}\\
e_{j} & \text{if }j\in\{p+1,...,p+n\},
\end{cases}
\]
(where $e_{j}$ is the vector of all zeros except for the $j$th element,
which is 1) we have}

\textcolor{black}{
\[
\widehat{\gamma}_{j}\leftarrow\begin{cases}
-1_{N\times1}^{T}Z._{j}+(\tilde{Z}\tilde{\gamma})^{T}Z._{j}+\gamma_{j}Z._{j}^{T}Z._{j} & \text{if }j=0\\
\frac{S\left((\tilde{Z}\tilde{\gamma})^{T}Z._{j}-1_{N\times1}^{T}Z._{j}-(\lambda_{G}/2)\tilde{\gamma}^{T}(G'_{\neq j}.)._{j},\ \lambda_{1}/2\right)}{Z._{j}^{T}Z._{j}+\lambda_{G}G'_{jj}} & \text{if }j\in\{1,...,p\}\\
\frac{H\left((\tilde{Z}\tilde{\gamma})^{T}e_{j}-1_{N\times1}^{T}e_{j},\delta\right)}{e_{j}^{T}e_{j}} & \text{if }j\in\{p+1,...,p+n\},
\end{cases}
\]
 yielding update:
\[
\hat{\gamma}_{j}\leftarrow\begin{cases}
\tilde{\gamma}^{T}\tilde{Z}1_{N\times1}+N(\gamma_{j}-1) & \text{if }j=0\\
\\
\frac{S\left((\tilde{Z}^{T}\tilde{\gamma}-1_{N\times1})^{T}X._{j}-(\lambda_{G}/2)\tilde{\beta}^{T}(G{}_{\neq j}.)._{j},\ \lambda_{1}/2\right)}{X._{j}^{T}X._{j}+\lambda_{G}G{}_{jj}} & \text{if }j\in\{1,...,p\}\\
\\
H\left((\tilde{Z}\tilde{\gamma})_{j}-1,\delta\right) & \text{if }j\in\{p+1,...,p+n\},
\end{cases}
\]
}where \textcolor{black}{where $S(x,\lambda)$ is the element-wise
soft-thresholding operator in equation \eqref{eq:soft-thresh} and}
\begin{equation}
H(x,\delta)=\begin{cases}
x-\delta & \text{if }x<1\\
x & \text{otherwise}.
\end{cases}\label{eq: H}
\end{equation}

\subsection*{\textcolor{black}{\medskip{}
Parameter grid used in cross-validation\medskip{}
}}

\begin{singlespace}
\textcolor{black}{Parameters $\{\lambda_{1},G,\lambda_{G},\delta,\lambda_{1}^{*}\}$
were taken over the following grid of values:
\begin{eqnarray*}
\lambda_{1} & \in & \{10,11,\ldots,99\}\\
G & \in & \left\{ L,\ L+\eta I,\ L+10^{2}\eta I,\ L+10^{3}\eta I,\ L+10^{4}\eta I\right\} \ \text{where }\eta=1/\lambda_{G}\ \text{for }\lambda_{G}>0\ \text{and }1\ \text{otherwise}\\
\lambda_{G} & \in & \{0,10^{1},10^{2},10^{3},10^{4},10^{5}\}\\
\delta & \in & \{0.2,0.3,0.4,0.5,0.6,0.7,1,2,10,100\}\\
\lambda_{1}^{*} & \in & \{1,10^{-1},10^{-2}\}.
\end{eqnarray*}
The linear SVM was fit over parameters 
\begin{equation}
C\in\{10{}^{-6},10^{-5},10^{-4},10^{-3},10^{-2},10^{-1},10^{-0},2,3,4,5,6,7,10^{1},10^{2},10^{3}\}.\label{eq:16}
\end{equation}
\medskip{}
}
\end{singlespace}

\section*{\textcolor{black}{References}}

\begin{singlespace}
\textcolor{black}{\bibliographystyle{elsarticle-harv}
\bibliography{NI_graphnet}
}\end{singlespace}

\end{document}